\documentclass[twocolumn, superscriptaddress, aps, prl, showpacs]{revtex4}

\usepackage{amsmath, amssymb, amsfonts}
\usepackage{graphicx}
\usepackage{sistyle}
\usepackage{bm}
\usepackage[usenames,dvipsnames]{xcolor}
\usepackage{hyperref}
\usepackage{nicefrac}
\usepackage{color}
\usepackage{multirow}

\usepackage{subfigure}

\graphicspath{{./pdf/}}

\newcommand{\abs}[1]{\left| #1 \right|}

\newcommand{\brc}[1]{\left\{ #1 \right\}}
\newcommand{\brck}[1]{\left[ #1 \right]}
\newcommand{\bra}[1]{\left\langle #1 \right|}
\newcommand{\ket}[1]{\left| #1 \right\rangle}
\newcommand{\braket}[2]{\left\langle {#1 {\left| \vphantom{#1 #2} \right.} #2} \right\rangle}

\newcommand{\paren}[1]{\left( #1 \right)}
\newcommand{\Tr}{\operatorname{Tr}}

\usepackage{array}
\newcolumntype{C}{>{\(}c<{\)}}

\begin{document}

\title{Characterization of high-dimensional entangled systems\\via mutually unbiased measurements}

\author{D.~Giovannini}
\thanks{d.giovannini.1@research.gla.ac.uk}
\affiliation{School of Physics and Astronomy, SUPA, University of Glasgow, Glasgow G12 8QQ, United Kingdom}
\author{J.~Romero}
\affiliation{School of Physics and Astronomy, SUPA, University of Glasgow, Glasgow G12 8QQ, United Kingdom}
\affiliation{Department of Physics, SUPA, University of Strathclyde, Glasgow G4 ONG, United Kingdom}
\author{J.~Leach}
\affiliation{School of Engineering and Physical Sciences, SUPA, Heriot-Watt University, Edinburgh EH14 4AS, United Kingdom}
\author{A.~Dudley}
\affiliation{CSIR National Laser Centre, PO Box 395, Pretoria 0001, South Africa}
\author{A.~Forbes}
\affiliation{CSIR National Laser Centre, PO Box 395, Pretoria 0001, South Africa}
\affiliation{School of Physics, University of KwaZulu-Natal, Private Bag X54001, Durban 4000, South Africa}
\author{M.~J.~Padgett}
\affiliation{School of Physics and Astronomy, SUPA, University of Glasgow, Glasgow G12 8QQ, United Kingdom}

\begin{abstract}
Mutually unbiased bases (MUBs) play a key role in many protocols in quantum science, such as quantum key distribution. However, defining MUBs for arbitrary high-dimensional systems is theoretically difficult, and measurements in such bases can be hard to implement.
We show experimentally that efficient quantum state reconstruction of a high-dimensional multi-partite quantum system can be performed by considering only the MUBs of the individual parts.
The state spaces of the individual subsystems are always smaller than the state space of the composite system. Thus, the benefit of this method is that MUBs need to be defined for the small Hilbert spaces of the subsystems rather than for the large space of the overall system. This becomes especially relevant where the definition or measurement of MUBs for the overall system is challenging. We illustrate this approach by implementing measurements for a high-dimensional system consisting of two photons entangled in the orbital angular momentum (OAM) degree of freedom, and we reconstruct the state of this system for dimensions of the individual photons from \(d=2\) to \(5\).
\end{abstract}

% 42.50.Tx: Optical angular momentum and its quantum aspects
% 03.65.Wj: State reconstruction, quantum tomography
% 03.65.Ud: Entanglement and quantum nonlocality
% 03.67.Dd: Quantum cryptography and communication security

\maketitle

Mutually unbiased bases (MUBs) \cite{Wootters:1989, Ivanovic:1981} are a key concept in quantum science, as they are intimately related to the nature of quantum information \cite{Wehner:2010, Barnett:2009, Durt:2010a}. Measurements made in one of a set of MUBs provide no information about the state if this was prepared in another basis from the same set. In quantum mechanics, the amount of information that can be extracted from a physical system is fundamentally limited by the uncertainty relations \cite{Wehner:2010, Barnett:2009}. In this context, MUBs acquire a fundamental relevance because they serve as a test bed with which one can explore general uncertainty relations and, ultimately, complementarity \cite{Durt:2010a}. Some important questions related to MUBs remain open \cite{Durt:2010a, Wehner:2010}: what is the number of MUBs for an arbitrary dimension \(d\), and why is mutual unbiasedness not enough to guarantee a strong uncertainty relation? While we do not seek to answer these questions, we provide an accessible experimental platform for exploring these problems by demonstrating measurements in complete sets of MUBs.

Many quantum information protocols depend upon the use of MUBs. For example, quantum key distribution (QKD) relies on the fact that measurements in one basis preclude knowledge of the state in any of the others \cite{Bennett:1984, Jun-Lin:2010, Malik:2012a}. In addition, MUBs play an important role in the reconstruction of quantum states \cite{Wootters:1989, Filippov:2011, Fernandez-Perez:2011}, where they have been successfully used to enable the optimal reconstruction of entangled states of polarization \cite{Adamson:2010} and single-photon linear momentum states \cite{Lima:2011}.

It is known that a Hilbert space of dimension \(D\) will have at most \(D+1\) MUBs \cite{Ivanovic:1981, Wootters:1989, Bandyopadhyay:2002}. In 1989, Wootters showed that if one can find \(D+1\) mutually unbiased bases in dimension \(D\), these bases provide a set of measurements that can be used to optimally determine the density matrix of a \(D\)-dimensional system \cite{Wootters:1989}. However, this approach rapidly breaks down for large \(D\) for two reasons: first, defining MUBs in high dimensions becomes increasingly difficult \cite{Durt:2010a, Brierley:2010a}, and second, performing the measurements in a complete high-dimensional set of MUBs becomes experimentally challenging \cite{Adamson:2010, Nagali:2010a}. This is especially relevant for multi-level multi-particle systems, where the dimension of the overall system scales as \(D=d^N\), with \(d\) the dimension of the Hilbert spaces of the \(N\) individual particles.

\begin{figure}[t]
\includegraphics[width=0.7\linewidth]{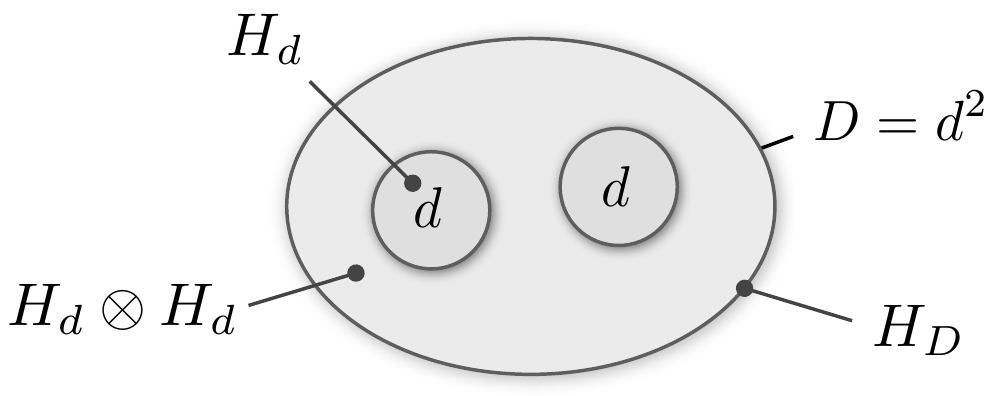}
\caption{\label{BipartiteSystem}Illustration of the state spaces of a bipartite system, where the system has dimension \(D\) and each subsystem \(d\). Adamson and Steinberg \cite{Adamson:2010} performed measurements in the Hilbert space \(H_D\) of the composite system, while we perform joint local measurements in the spaces \(H_d\) of the individual subsystems (for \(d=2\) to \(5\)).}
\end{figure}

We show experimentally that the alternative approach of performing local measurements in the MUBs of the single particles of a multi-particle system still allows a complete reconstruction of the overall density matrix with a minimum number of measurements \cite{Thew:2002a}. The significant benefit of our procedure is that it only requires the definition of MUBs in a Hilbert space of size \(d=D^{1/N}\); see Fig.~\ref{BipartiteSystem}. We illustrate this approach in the case of a photonic implementation of a bipartite multi-level entangled system (\(d=\sqrt{D}\)) using the orbital angular momentum of light.

In addition to the spin angular momentum, associated with polarization, light can also carry orbital angular momentum (OAM) \cite{Yao:2011a}. The OAM of light is associated with phase structures of the form \(e^{i\ell\phi}\), where \(\ell\hbar\) is the OAM carried by each photon and \(\phi\) the azimuthal angle \cite{Allen:1992}. The unbounded Hilbert space of OAM is one example of a scalable high-dimensional resource that can be used for quantum information science \cite{Mair:2001, Leach:2010, Groblacher:2006, Salakhutdinov:2012}. For example, the entanglement of high-dimensional states provides implementations of QKD that are more tolerant to eavesdropping and can improve the bit rate in other quantum communication protocols \cite{Bourennane:2002, Cerf:2002, Bechmann-Pasquinucci:2000, Walborn:2006, Dixon:2012, Gruneisen:2012}.

One of the advantages of OAM is the ability to access \(d\)-dimensional subspaces \cite{Dada:2011}, for each of which we can define all existing MUBs \cite{Gruneisen:2008}. In this work, we implement measurements in high-dimensional MUBs within the OAM degree of freedom, and we show that the MUBs corresponding to \(d\)-dimensional subspaces are readily accessible with simple laboratory procedures. Furthermore, we show that measurements in MUBs of these subspaces can be used for the complete tomographic reconstruction of multipartite entangled systems with the minimum number of measurements. We produce entangled photon pairs by means of spontaneous parametric down-conversion (SPDC) that we then measure in full sets of \(d+1\) MUBs for OAM, for dimensions ranging from \(d=2\) to \(5\). The states belonging to the MUBs are defined as superpositions of Laguerre-Gaussian (LG) modes.

{\it Theory:} Consider two operators in a \(d\)-dimensional Hilbert space with orthonormal spectral decompositions. These operators, and their basis states, are said to be mutually unbiased \cite{Ivanovic:1981, Wootters:1989} if
\begin{equation}
\abs{\braket{\psi_{m,i}}{\psi_{n,j}}}^2 =
\begin{cases}
1/d & \text{for } m\neq n \\
\delta_{ij} & \text{for } m=n
\end{cases}
\end{equation}
for all \(i\) and \(j\). The indices \(i\) and \(j\) correspond to the basis states, and \(m\) and \(n\) indicate any two bases. Operators that are quantum-mechanical observables are sometimes called mutually complementary, or maximally noncommutative \cite{Schwinger:1960}. This is because, given any eigenstate of one, the eigenvalue resulting from a measurement of the other is completely undetermined. In other words, the state of a system described in one MUB provides no information about the state in another. It is known that the number of MUBs in dimension \(d\) cannot exceed \(d+1\) \cite{Wootters:1989, Durt:2010a}, and it is exactly \(d+1\) if \(d\) is prime or a prime power \cite{Wootters:1989, Klappenecker:2004}.

The simplest set of mutually unbiased observables can be found in dimension \(d=2\). For example, in the two-dimensional Hilbert space of polarization, the bases of horizontal/vertical, diagonal/anti-diagonal and left/right circular polarizations provide a set of three MUBs. Two states belonging to the same basis are orthonormal, while the square of the inner product of any two states belonging to different bases is always \(1/2\). Equivalent mutually unbiased states can be implemented using other two-dimensional state spaces, e.g. a subspace of OAM.

In our work, we choose to investigate the OAM degree of freedom of single photons. A general single-photon state in a \(d\)-dimensional subspace can be described by an orthonormal basis set of OAM modes \(\ket{\ell}\) as \(\ket{\psi} = \sum_{\brc{\ell}} c_\ell \ket{\ell}\).
The complex coefficients \(c_\ell\) are subject to the normalization condition \(\sum c_\ell^2 = 1\). Defining MUBs in a general \(d\)-dimensional space is a difficult problem \cite{Brierley:2010a}; however, for a number of low-dimensional cases, it is possible to find complete sets of MUBs using simple procedures \cite{Brierley:2010}. For these cases, which include the dimensions 2 to 5, the states \(\brc{\ket{\ell}}\) can be chosen to be one of the MUBs. The states belonging to the remaining \(d\) MUBs are found to be superpositions of the basis states with coefficients of equal magnitude \(\abs{c_\ell}=1/\sqrt{d}\) but differing phases.

In general, it is possible for a system to include more than one particle. If one considers a \(d\)-dimensional state space for each particle, the dimension \(D\) of a system of \(N\) particles will be \(D=d^N\). Such a system will be unambiguously specified by its density matrix \(\rho\), a positive-semidefinite unit-trace Hermitian operator that includes \(d^{2N}-1\) independent real parameters (\(d^4-1\) for a bipartite system).

MUBs play an important role in quantum state tomography (QST) \cite{Wootters:1989, Klimov:2008}, the process of determining the density matrix of an unknown quantum system \cite{James:2001, Langford:2004, Altepeter:2005}. One approach to tomography is to perform measurements in the MUBs of the \(D\)-dimensional state space of the composite system \cite{Wootters:1989}. However, such measurements are very challenging as they require the definition of MUBs for Hilbert spaces of very high dimension and can require the implementation of entangled observables \cite{Adamson:2010}. Our approach is simpler as we use the MUBs of the state spaces of the single particles.

Let us consider for simplicity a bipartite system. An overcomplete set of measurements for the reconstruction of the \(D\)-dimensional system is provided by the pairwise combinations of all single-particle MUB states. The total number of independent measurements for this approach is equal to \((d(d+1))^2\), which is always greater than \(d^{4}-1\). We propose another suitable set of measurements, given by pairwise combinations of states from an appropriate subset of the overcomplete set. This subset contains all states in one MUB and all but one state in each of the remaining \(d\) MUBs. It can be shown that the conditions for the completeness of a set of tomographic measurements \cite{Altepeter:2005} are satisfied by this reconstruction strategy (see supplemental material).

This approach gives exactly the \(d^4\) independent measurements that can then be used for a tomographically complete reconstruction of the \(D\)-dimensional system. The number of measurements in our method scales favourably with the dimension of the system if compared with other methods (see supplemental material).

{\it Experimental methods:} A \(\SI{3}{mm}\)-thick \(\beta\)-barium borate (BBO) non-linear crystal cut for type-I collinear SPDC is pumped by a collimated \(\SI{1}{W}\) UV laser to produce frequency-degenerate entangled photon pairs at \(\SI{710}{nm}\). The co-propagating signal and idler photons are separated by a non-polarizing beam splitter and redirected to spatial light modulators (SLMs), onto which the output face of the crystal is imaged by a \(2\times\) telescope. In order for the crystal to produce two-photon states entangled over a wider range of OAM modes, we tune the phase-matching conditions of the BBO crystal to increase the OAM spectrum of the down-converted state \cite{Romero:2012b}. The SLMs act as reconfigurable computer-generated holograms (CGHs) that allow us to measure any arbitrary superposition of OAM modes. The SLMs are used to modulate the phase and introduce a spatially dependent attenuation to discard light into the zero diffraction order, allowing the manipulation of the complex amplitude of the incoming light \cite{Arrizon:2007, Davis:1999, Gruneisen:2008}.

We pump the crystal with a plane phase front. In order to observe correlations in all bases (instead of anti-correlations), the hologram displayed in one of the two detection arms is phase-conjugate with respect to the other \cite{Mair:2001}. The projected Gaussian mode is then imaged onto a single-mode fibre (SMFs) that is coupled to a single-photon photodiode detector. The detectors' outputs are routed to coincidence-counting electronics with a timing window of \(\SI{10}{ns}\). Narrow-band, \(\SI{10}{nm}\) interference filters are placed in front of the detectors to ensure that the frequency spread of the detected down-converted fields is small compared to the central frequencies.

\begin{figure}[t]
\includegraphics[width=\linewidth]{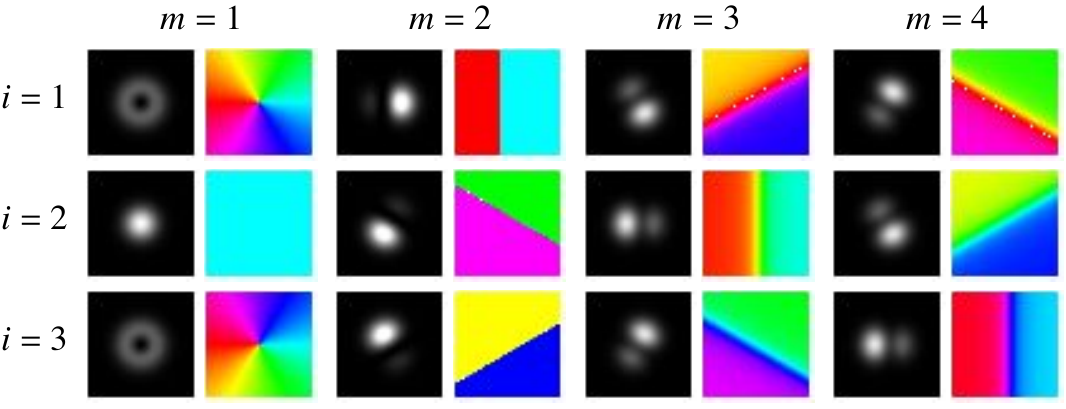}
\caption{\label{MU3Modes}Mutually unbiased modes \(i\) for each of the 4 bases \(m\) in \(d=3\). The greyscale images represent the intensity; the colour images represent the phase. The first basis, \(m=1\), corresponds to Laguerre-Gaussian modes with OAM ranging from \(\ell = -1\) to \(+1\).}
\end{figure}

The combination of the two SLMs, single-mode fibres and coincidence-counting electronics allows us to perform projective measurements on the entangled state of photons \(A\) and \(B\) described by the operators
\begin{equation}
\Pi_{m, i; n, j} = \ket{\psi_{m,i}}_A\ket{\psi_{n,j}}_B^\ast \bra{\psi_{m,i}}_A\bra{\psi_{n,j}}_B^\ast.
\end{equation}
Here, the single-photon states \(\ket{\psi}_A\) and \(\ket{\psi}_B\) belong to MUBs in \(d\) dimensions and are given by
\begin{equation}
\ket{\psi_{m,i}} = \sum_{\brc{\ell}} c_{m,i,\ell} \, \ket{\ell},
\end{equation}
where \(c_{m,i,\ell}\) is a complex coefficient. The indices \(m\) and \(n\), which correspond to the basis indices, range from \(1\) to \(d+1\); the indices \(i\) and \(j\), which represent a state within a basis, range from \(1\) to \(d\). For each dimension \(d\), we choose one set of OAM states \(\brc{\ket{\ell}}\).
The OAM values used are \(\brc{\ell} = \brc{-2,\,+2}\) for \(d=2\), \(\brc{-2, \, -1,\, +1,\, +2}\) for \(d=4\), and \(\brc{-\lfloor d/2 \rfloor, \, \dots, \, +\lfloor d/2 \rfloor}\) for \(d=3\) and \(5\).
For each \(d\), we take the basis corresponding to \(m=1\) to be the orthonormal basis given above; the remaining bases are composed of superpositions of the \(m=1\) states with appropriate complex coefficients; see Fig.~\ref{MU3Modes}. For the dimensions considered, the magnitude of these complex coefficients is \(1/\sqrt{d}\) for all \(i\) and \(\ell\).

To determine the phase terms \(c_{m,i,\ell}\) that define the MUBs (for \(m = 2\) to \(d+1\)), we use the methods outlined by Refs.~\cite{Brierley:2010, Brierley:2009a}. The coefficients are given by the mutually unbiased vectors derived from \(d \times d\) dephased Hadamard matrices. These matrices are unique for \(d=2, 3,\,4\) and \(5\). For \(d=2\), the MUBs obtained are the familiar set of bases that one usually associates with polarization states. Consequently, the two-dimensional MUBs for OAM \cite{Padgett:1999} are the analogue of those for polarization \cite{Bruss:1998}. All the modes used for \(d=3\) are shown in Fig.~\ref{MU3Modes}.

\begin{figure}[t]
\includegraphics[width=0.49\linewidth]{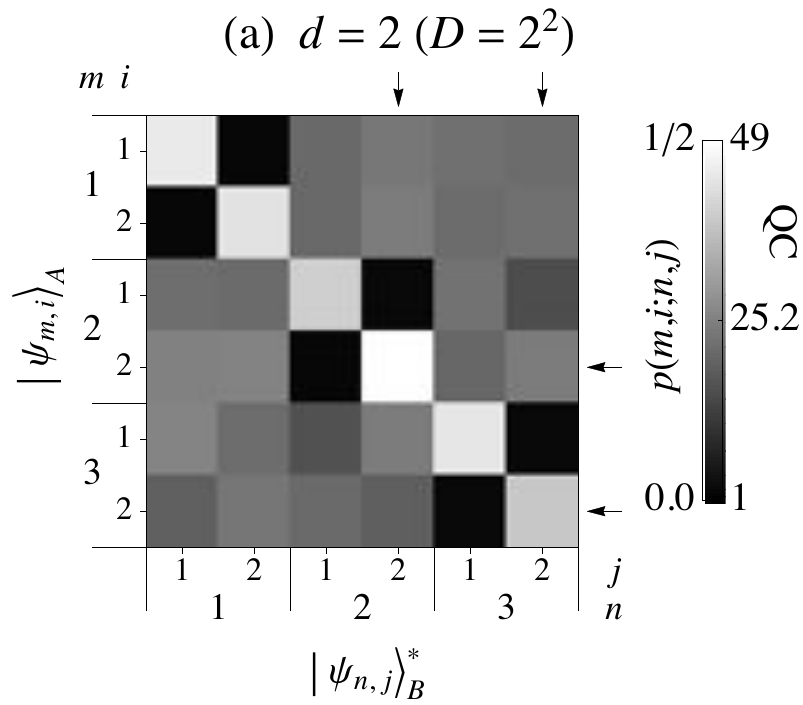} \hfill
\includegraphics[width=0.49\linewidth]{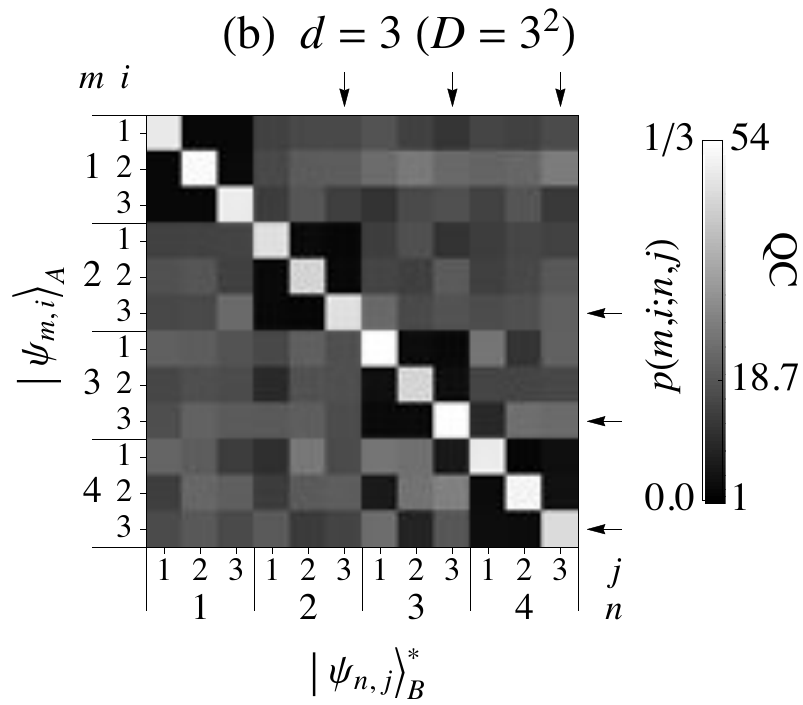} \\[5pt]
\includegraphics[width=0.49\linewidth]{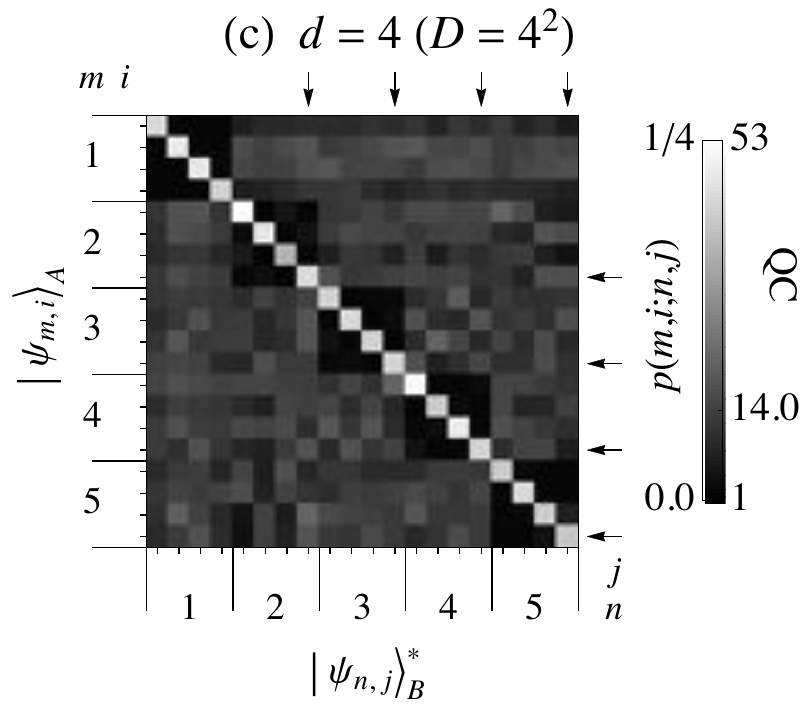} \hfill
\includegraphics[width=0.49\linewidth]{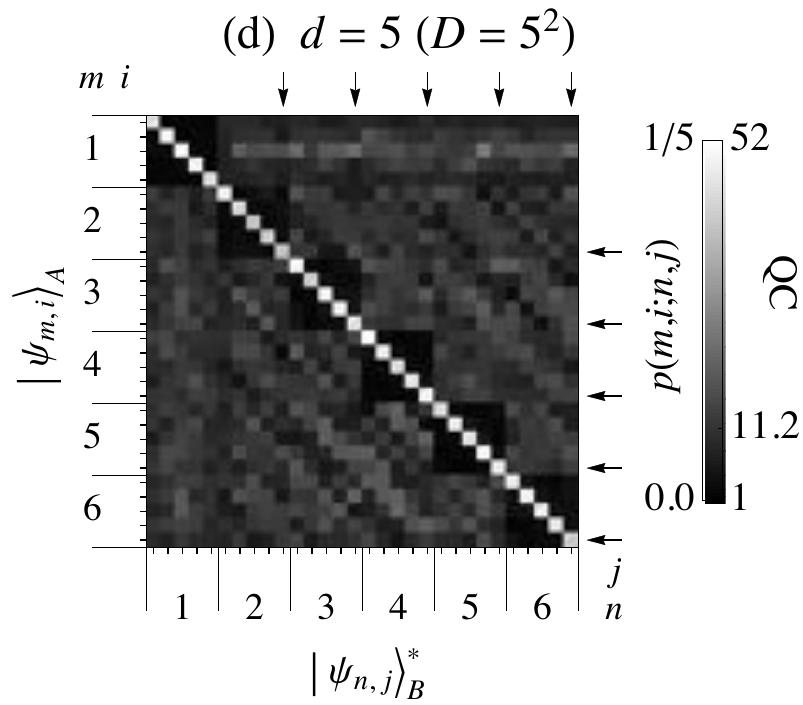}
\caption{\label{CorrelationsMatrices}Joint probabilities of detecting photon A in state \(\ket{\psi_{m, i}}_A\) and photon B in state \(\ket{\psi_{n, j}}_B\). The results are normalized such that the sum of the joint detection probabilities for measurements in any two bases \(m\) and \(n\) are unity. Therefore, the probabilities represented by the leading diagonal are expected to be \(1/d\), and all probabilities for \(m \neq n\) are expected to be \(1/d^2\). We also display the quantum contrast QC, which is given by the ratio of the measured coincidence rate to that of the expected accidental coincidences. The arrows indicate the rows and columns of measurements not required for the complete tomographic reconstruction of the density matrix.}
\end{figure}

An overcomplete set of measurements is obtained by scanning through all possible values of \(m\) and \(i\), for photon \(A\), and \(n\) and \(j\), for photon \(B\). For every combination of \(m, n, i\) and \(j\), we record the coincidence counts and both the single channel counts resulting from the projective measurement. From this set of data we extract the tomographically complete set of measurements previously described. These count rates are converted to detection probabilities through the following relationship:
\begin{equation}
p_k = \frac{d^2}{\sum{C_k}} \frac{C_k - U_k}{U_k},
\end{equation}
where the index \(k\) corresponds to a unique choice of measurement settings \(m, n, i\) and \(j\), \(C_k\) is the coincidence count rate and \(U_k\) is the anticipated uncorrelated coincidence rate, which is estimated by taking the product of the single-channel count rates and the gate time (Fig.~\ref{CorrelationsMatrices}). The normalization approach that we take accounts for different hologram efficiencies for different modes (see supplemental material).

The task of the fitting procedure is to find the optimal density matrix \(\rho\) of the \(D\)-dimensional system that best reproduces the experimental results. The parameters of the density matrix are established through numerical minimization of the Pearson's cumulative test statistic \cite{Opatrny:1997, Banaszek:1999}
\begin{equation}
\chi^2 = \sum_{k=1}^{d^{4}} \frac{(p_k-p_k^\prime)^2}{p_k^\prime},
\end{equation}
where \( p_k \) are the probabilities from the experiment, and \(p_k^\prime = \Tr[\rho \, \Pi_k]\) are those predicted from the reconstructed density matrix.

\begin{figure}[t]
\includegraphics[width=0.49\linewidth]{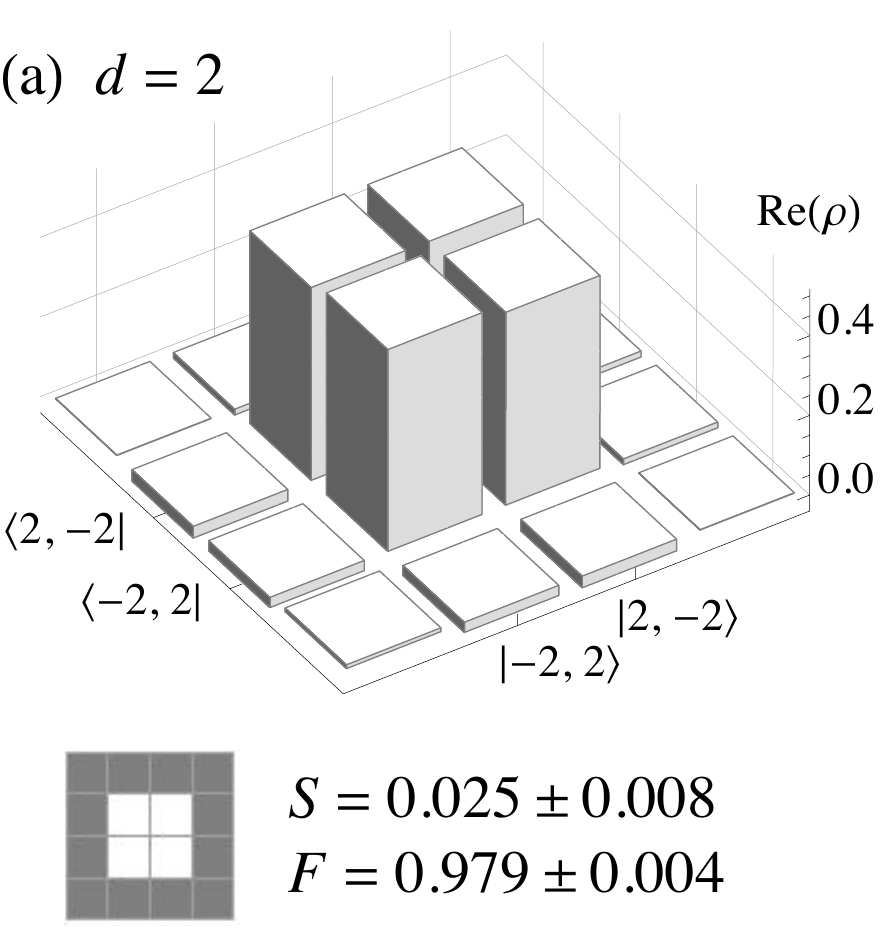} \hfill
\includegraphics[width=0.49\linewidth]{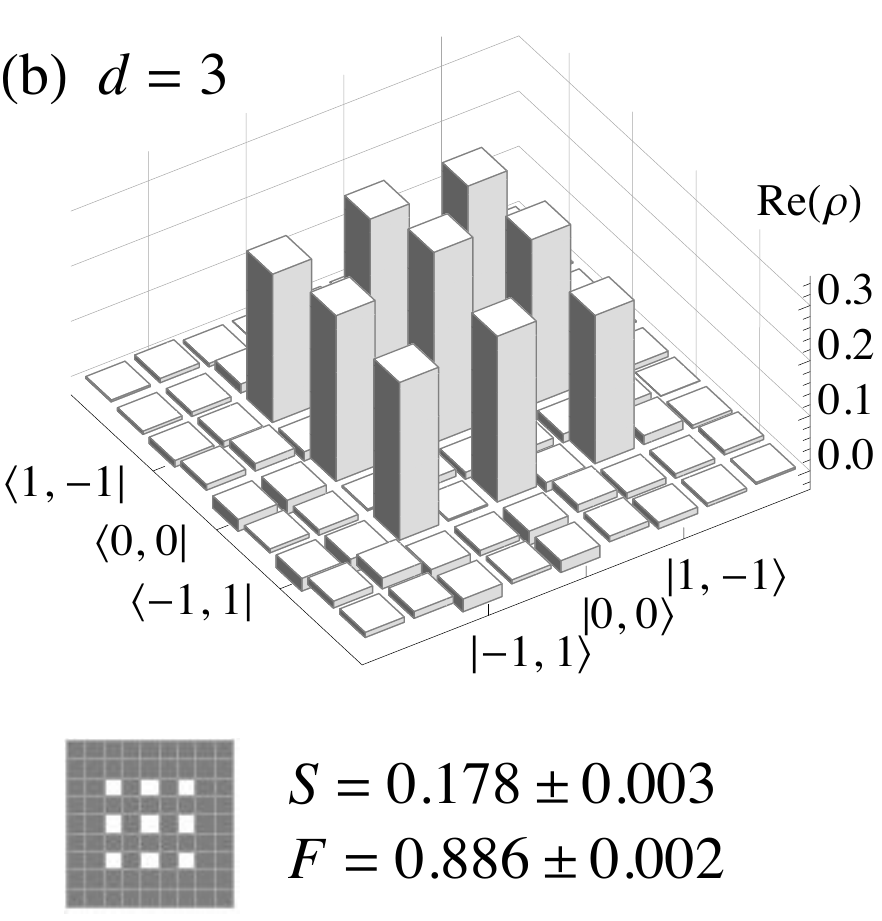} \\
\includegraphics[width=0.49\linewidth]{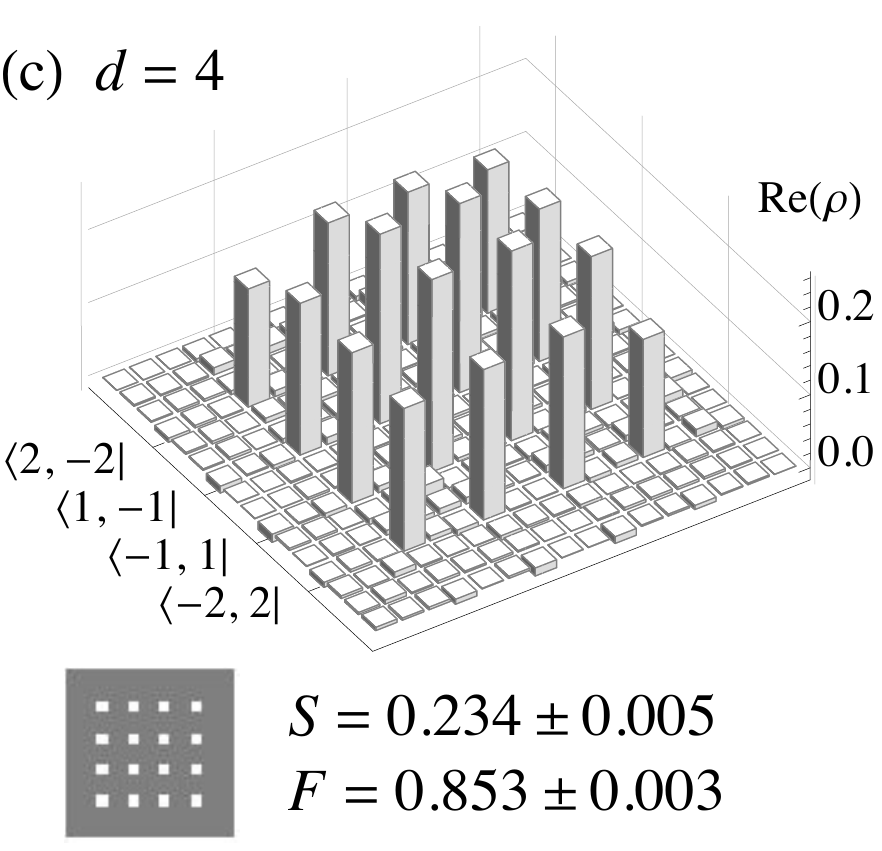} \hfill
\includegraphics[width=0.49\linewidth]{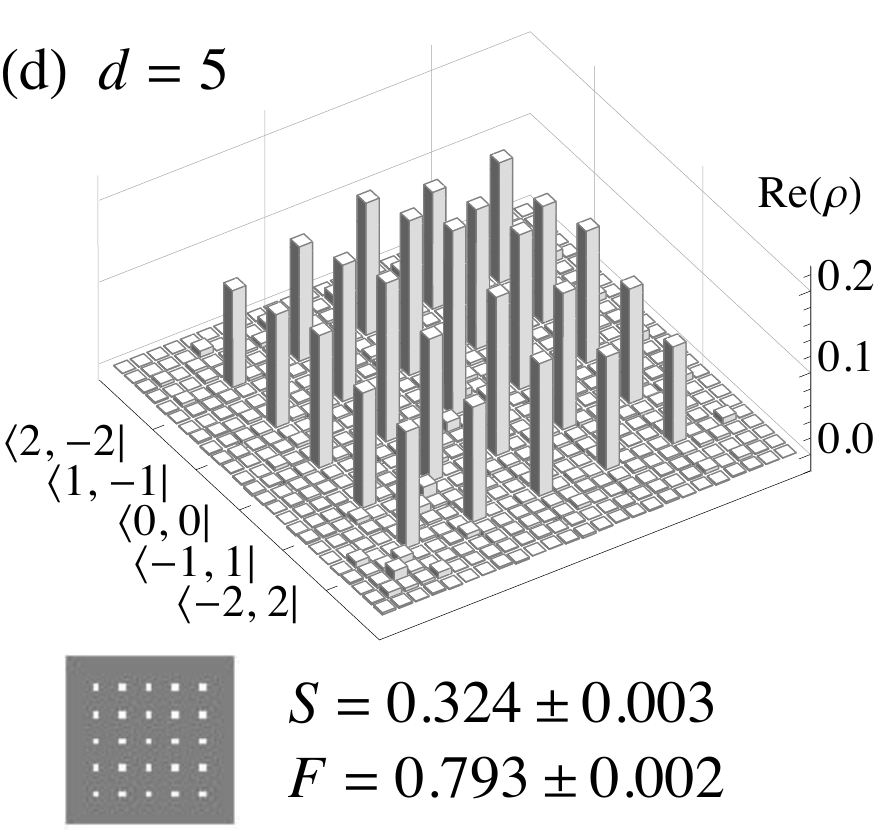}
\caption{\label{ReconstructedMatrices}Results of tomographic reconstructions using a complete set of single-photon mutually unbiased bases measurements. The real parts of the reconstructed density matrices \(\rho\) are shown. Imaginary parts are less than \(0.076\) for \(d=2\), \(0.059\) for \(d=3\), and \(0.050\) for \(d=5,6\). Also shown are the linear entropy \(S\) and fidelity \(F\) for the reconstructed density matrices. Insets: real parts of the theoretical density matrices for the maximally entangled states.}
\end{figure}

The reconstructed density matrices for dimensions \(2, 3, 4\) and \(5\) are shown in Fig.~\ref{ReconstructedMatrices}. For each reconstructed density matrix \(\rho\), we calculate the linear entropy \(S=1-\Tr(\rho^2)\) and the fidelity \(F=\Tr[\sqrt{\sqrt{\sigma}\rho\sqrt{\sigma}}]^2\), where \(\sigma \) is the \(D\)-dimensional maximally entangled density matrix associated with arbitrarily large spiral bandwidth \cite{Romero:2012b} and perfect detection. The uncertainties were calculated by repeating the reconstruction process for statistically equivalent copies of the original experimental data sets, each obtained by adding Poissonian fluctuations to the measured counts.

The reconstructed density matrices have low entropies, indicating pure states, and very high fidelities with respect to the maximally entangled state. Due to the finite spiral bandwidth of our generated state \cite{Torres:2003a, Romero:2012b} and limitations in our measurement system, one would anticipate the fidelities to decrease and the entropies to increase as the dimension increases. Indeed, we observe this characteristic in our results.

For comparison, we also implemented the approach described in \cite{Agnew:2011}. We find comparable entropies and fidelities whichever approach is used (see supplemental material). However, our method requires significantly fewer measurements. For example, for \(d=5\), the number of measurements required is \(d^4=625\) compared to \(2025\) for the procedure outlined in Ref.~\cite{Agnew:2011}. Both methods rely on projective measurements in appropriate superpositions of the basis states in the dimension of choice. Neither is more experimentally demanding, as they can both be performed using the same setup and only differ in the choice of projection states.

The MUBs reconstruction method is applied here to almost maximally entangled states. The density matrices of maximally entangled states have low rank, \(r<D\), and could thus be efficiently reconstructed through compressed sensing \cite{Gross:2010, Liu:2012}. In the general case, however, a complete quantum state reconstruction by means of appropriately selected projection operators may be more appropriate and produce results with higher fidelity. 

{\it Conclusions:} In this work, we have demonstrated single-photon measurements for MUBs in the OAM degree of freedom and shown how these measurements can be used for efficient quantum state reconstruction. The procedure of measuring combinations of all single-photon states in one basis and all but one state in the remaining bases gives a minimal complete set of tomographic measurements. This experimental method can be readily applied to multi-level multi-partite systems.

The OAM degree of freedom is becoming an important resource for quantum information science. Therefore, the ability to measure states in MUBs is an important step for quantum protocols implemented in this degree of freedom. Measuring MUBs in high-dimensional spaces is not just of practical importance for QKD protocols, but it can also provide important insight into the nature of information in physical systems.

{\it Acknowledgements:} We thank Adetunmise Dada and Daniel Tasca for useful discussions. This work was supported by EPSRC. MJP thanks the Royal Society.

\appendix

\section{Supplemental material}

\subsection{Normalization of probabilities}

After recording the coincidence count rates \(C_k\) for each choice of \(n, i, m\) and \(j\), and the single-channel count rates \(A_k\) and \(B_k\), we convert the count rates to detection probabilities through
\begin{equation}
\label{ExpProbabilities}
p_k = \Upsilon \paren{\frac{C_k - A_k B_k \Delta t}{A_k B_k \Delta t}},
\end{equation}
where \(\Delta t\) is the gate time of our coincidence-counting electronics and \(\Upsilon\) an appropriate normalization factor. The term \(A_k B_k \Delta t\) corresponds to the uncorrelated accidental count rate \(U_k\).

\begin{figure}[t]
\subfigure[Overcomplete QST]{\label{MeasMatrixOC}\includegraphics[width=0.54\linewidth]{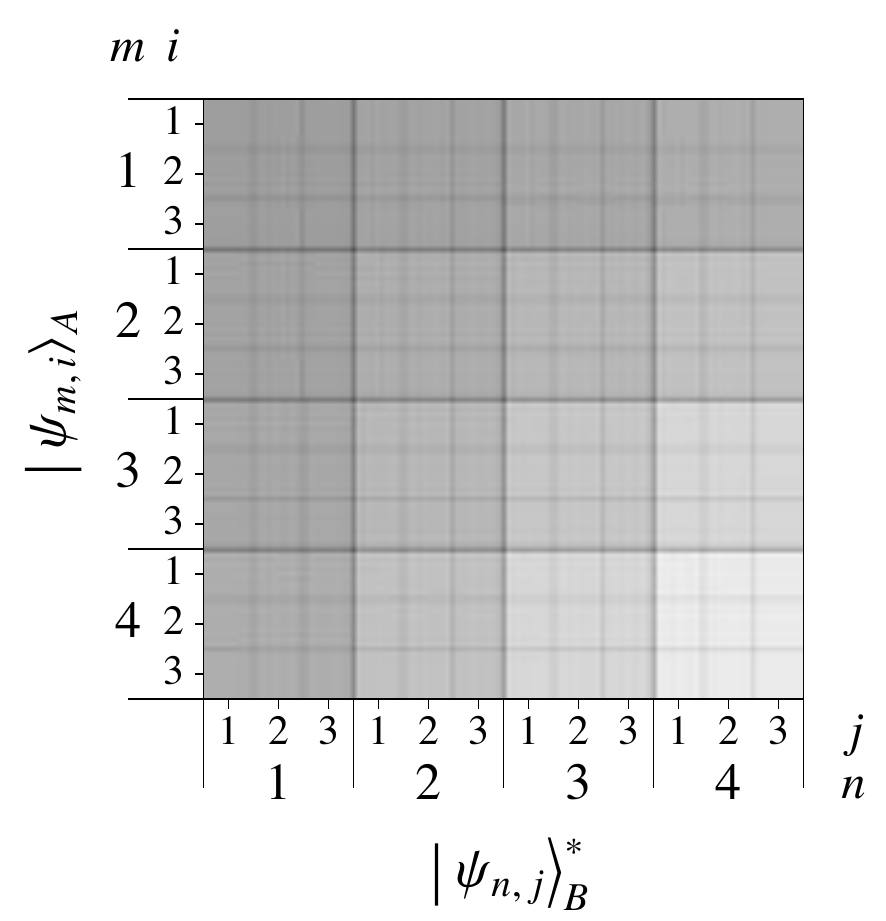}} \hfill
\subfigure[Complete QST]{\label{MeasMatrixC}\includegraphics[width=0.44\linewidth]{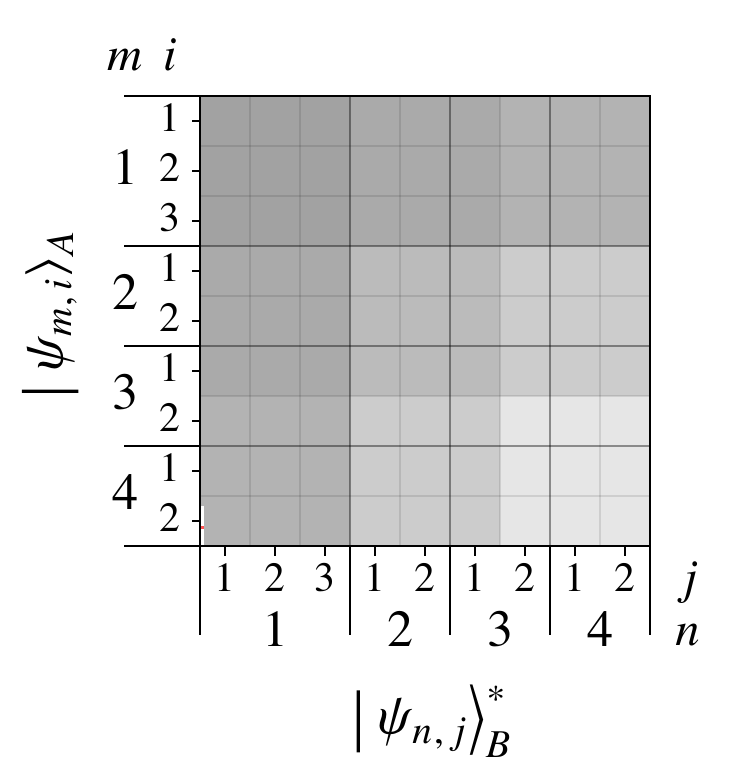}}
\caption{Sets of measurements for \subref{MeasMatrixOC} overcomplete state tomography and \subref{MeasMatrixC} complete state tomography, for \(d=3\) (\(D=3^2\)). The shaded areas indicate quadrants of size \(d \times d\), for each of which we expect \(\sum_k p_k=1\).}
\end{figure}

The normalization factor
\begin{equation}
\Upsilon = Q/\sum_{k=1}^{d^2 Q} C_k
\end{equation}
depends on the type of tomographic reconstruction performed. The factor \(Q\) indicates the number of \(d \times d\) quadrants in the correlations matrix for the set of measurements of choice. The product \(d^2 Q\) corresponds to the total number of independent measurements. For an overcomplete tomography, where we set \(\sum_k^{d^2} p_k = 1\) for any given choice of \(m\) and \(n\), \(Q=(d+1)^2\); see Fig.~\ref{MeasMatrixOC}. For a tomographically complete reconstruction that uses the presented subset of MUBs measurements, \(Q=\brck{1+(d-1)}^2=d^2\); see Fig.~\ref{MeasMatrixC}.

\subsection{Completeness of tomographic reconstruction}

One can express the density matrix \(\rho\) as a linear combination of a complete basis of \(d^2 \times d^2\) matrices \(\Gamma_\mu\) with complex coefficients \(\gamma_\mu\) \cite{Thew:2002a}:
\begin{equation}
\rho = \frac{\Gamma_0}{D}+\sum_{\mu=1}^{D^2-1} \gamma_\mu \, \Gamma_\mu,
\end{equation}
where \(D=d^2\) is the dimension of our bipartite system.
The basis matrices \(\Gamma_\mu\) have the following properties:
\begin{subequations}
\begin{equation}
\Tr(\Gamma_\mu \cdot \Gamma_\nu) = \delta_{\mu,\nu}
\end{equation}
\begin{equation}
\kappa = \sum_\mu \Gamma_\mu \Tr(\Gamma_\mu \cdot \kappa),
\end{equation}
\end{subequations}
where \(\kappa\) is any \(d^2 \times d^2\) matrix. A suitable set of Hermitian matrices \(\Gamma_\mu\) for the decomposition of \(\rho\) is given by the generalized Gell-Mann matrices for dimension \(D\).

A necessary and sufficient condition for the completeness of the set of tomographic states \(\brc{\ket{\psi_\mu}}\) (associated with the two-qudit observables \(\Pi_\mu\)) is given by the invertibility of the matrix
\begin{equation}
\label{MatrixB}
B_{\mu \nu} = \bra{\psi_\mu} \Gamma_\nu \ket{\psi_\mu}
\end{equation}
which allows us to express the complex coefficients \(\gamma_\mu\) in terms of probabilities \(p_\mu=\bra{\psi_\mu}\rho\ket{\psi_\mu}\) \cite{Altepeter:2005}:
\begin{equation}
\gamma_\mu = d^2 \sum_{\nu=1}^{d^4} (B^{-1})_{\mu\nu} \, p_\nu.
\end{equation}

Let us define the orthonormal set of basis vectors \(u_i\) in dimension \(d\), whose elements are given by \((u_i)_j = \delta_{ij}\). For a choice of two single-particle MUBs vectors for the qu\(d\)it subsystems \(A\) and \(B\)
\begin{subequations}
\begin{equation}
u_A = (a_1, \, \dots \, a_d)
\end{equation}
\begin{equation}
u_B = (b_1, \, \dots \, b_d),
\end{equation}
\end{subequations}
we can express the elements \(j=1, \, \dots \, D\) of the corresponding vector for the \(D\)-dimensional state space of the bipartite system as
\begin{equation}
(v_{AB})_j = a_\alpha b_\beta,
\end{equation}
where \((\alpha, \, \beta)\) are all pairwise permutations of indices \(\brc{1, \, \dots \, d}\). From the \(D\)-dimensional vectors \(v_{AB}\) we then define the states \(\ket{\psi_\mu}\) that describe the measurements on the composite system.

After calculating the states \(\ket{\psi_\mu}\) for the subset of MUBs measurements for complete tomography defined previously, we find the invertible matrix \(B\) through Eq.~\eqref{MatrixB}.

\begin{figure}[t]
\includegraphics[width=0.55\linewidth]{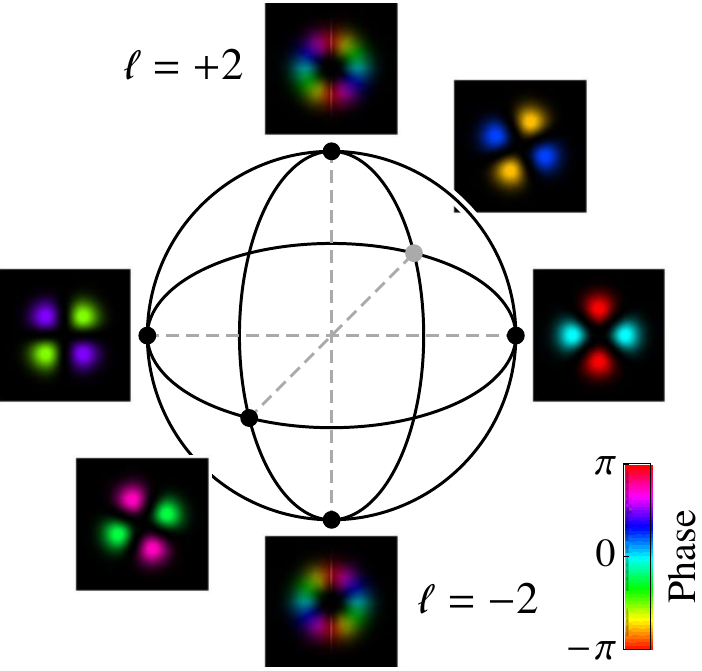}
\caption{\label{MU2Modes}Mutually unbiased modes for \(d=2\) in the \(\abs{\ell}=2\) OAM subspace. The brightness of the image corresponds to the intensity of the modes; the colour represents phase.}
\end{figure}

\subsection{Quantum state reconstruction via numerical optimization}

The number of joint measurements required to perform our reconstruction procedure is given by:
\begin{equation}
M_\text{MUBs} = d^2 + 2d^2(d-1)+d^2(d-1)^2 = d^4,
\end{equation}
which corresponds to the minimum number of parameters required to perform a complete quantum state tomography. The number of measurements required by the overcomplete quantum state reconstruction strategy outlined in Ref.~\cite{Agnew:2011} requires instead the following total number of measurements:
\begin{equation}
M_\text{QST} = \brck{4 \binom{d}{2}+d}^2.
\end{equation}

The numerical optimization to find the density matrix \(\rho\) that provides the best fit to the experimental probabilities from Eq.~\eqref{ExpProbabilities} is carried out by performing a random search over the parameter space of a complex left-triangular matrix \(T\) \cite{Altepeter:2005}, from which a physical guessed density matrix is derived:
\begin{equation}
\rho^\prime = T^\dag T/\Tr(T^\dag T).
\end{equation}
We reconstructed the states from \(d=2\) to \(5\) using both methods. A quantitative comparison of the results is shown in Tab.~\ref{ReconstructionResults}.

\begin{figure}[t]
\includegraphics[width=\linewidth]{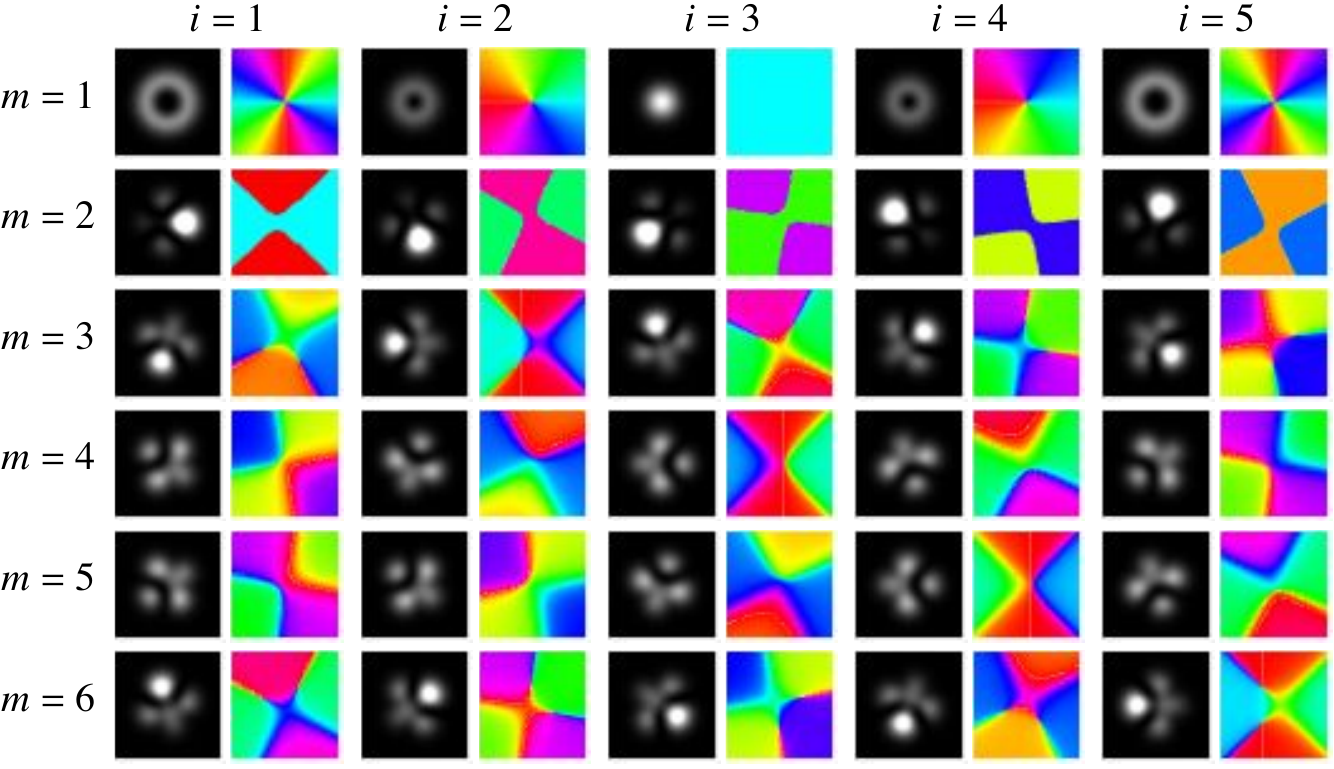}
\caption{\label{MU5Modes}Mutually unbiased modes \(i\) for each of the 6 bases \(m\) in \(d=5\). The greyscale images represent the intensity; the colour images represent the phase. The first basis, \(m=1\), corresponds to Laguerre-Gaussian modes with OAM ranging from \(\ell = -2\) to \(+2\).}
\end{figure}

\begin{table}
\caption{\label{ReconstructionResults}Linear entropy \(S\) and fidelity \(F\) (with respect to a maximally entangled density matrix) for density matrices reconstructed from overcomplete quantum state tomography (QST) and measurements in mutually unbiased bases (MUBs). \(M\) represents the number of measurements needed for the indicated reconstruction method.}
\begin{ruledtabular}
\begin{tabular}{ccccc}
& Method & \(M\) & \(S\) & \(F\) \\ \hline
\multirow{2}{*}{\(d=2\)}
& MUBs & 16 & \(0.025 \pm 0.008\) & \(0.979 \pm 0.004\) \\
& QST & 36 & \(0.070 \pm 0.007\) & \(0.958 \pm 0.004\) \\
\hline
\multirow{2}{*}{\(d=3\)}
& MUBs & 81 & \(0.178 \pm 0.003\) & \(0.886 \pm 0.002\) \\
& QST & 225 & \(0.179 \pm 0.005\) & \(0.893 \pm 0.003\) \\
\hline
\multirow{2}{*}{\(d=4\)}
& MUBs & 256 & \(0.234 \pm 0.005\) & \(0.853 \pm 0.003\) \\
& QST & 784 & \(0.281 \pm 0.009\) & \(0.818 \pm 0.006\) \\
\hline
\multirow{2}{*}{\(d=5\)}
& MUBs & 625 & \(0.324 \pm 0.003\) & \(0.793 \pm 0.002\) \\
& QST & 2025 & \(0.364 \pm 0.008\) & \(0.764 \pm 0.006\)
\end{tabular}
\end{ruledtabular}
\end{table}
\vspace{1em}

\subsection{Mutually unbiased vectors}

The complete sets of mutually unbiased vectors used in the quantum state tomography, for dimensions from \(d=2\) to \(5\), are reported in this section. For each vector \(v_{mi}\), \(m\) indicates the basis among the \(d+1\) available in dimension \(d\) and \(i\) the vector within the basis. Each vector provides the corresponding set of complex coefficients for the superposition of the basis modes of choice; see Fig.~\ref{MU2Modes} for \(d=2\) and Fig.~\ref{MU5Modes} for \(d=5\).

The experimental procedure and the reconstruction technique can be readily extended to higher dimensions. The existence of full sets of \(d+1\) MUBs has however only been proven for dimensions \(d\) that are prime numbers or powers of a prime. Finding MUBs in higher prime power dimensions, especially sets that may be suitable for practical implementations, remains challenging. It should also be noted that, despite MUBs being particularly advantageous to efficiently reconstruct the density matrix of an unknown state encoded in the spatial modes of a single photon, as \(d\) increases the complicated structures of the modes involved may negatively affect the detection efficiency.

\subsubsection{Coefficients for \(d=2\)}

\begin{ruledtabular}
\begin{tabular}{CCCC}
m & i & c_1 & c_2 \\ \hline
\multirow{2}{*}{1} & 1 & 1 & 0 \\
& 2 & 0 & 1 \\ \hline
\multirow{2}{*}{2} & 1 & \nicefrac{1}{\sqrt{2}} & \nicefrac{1}{\sqrt{2}} \\
& 2 & \nicefrac{1}{\sqrt{2}} & \nicefrac{-1}{\sqrt{2}} \\ \hline
\multirow{2}{*}{3} & 1 & \nicefrac{1}{\sqrt{2}} & \nicefrac{i}{\sqrt{2}} \\
& 2 & \nicefrac{1}{\sqrt{2}} & \nicefrac{-i}{\sqrt{2}}
\end{tabular}
\end{ruledtabular}
\vspace{1em}

\subsubsection{Coefficients for \(d=3\)}

\begin{ruledtabular}
\begin{tabular}{CCCCC}
m & i & c_1 & c_2 & c_3 \\ \hline
\multirow{3}{*}{1} & 1 & 1 & 0 & 0 \\
& 2 & 0 & 1 & 0 \\
& 3 & 0 & 0 & 1 \\ \hline
\multirow{3}{*}{2} & 1 & \nicefrac{1}{\sqrt{3}} & \nicefrac{1}{\sqrt{3}} & \nicefrac{1}{\sqrt{3}} \\
& 2 & \nicefrac{1}{\sqrt{3}} & \nicefrac{e^{\nicefrac{2 i \pi }{3}}}{\sqrt{3}} & \nicefrac{e^{\nicefrac{-2 i \pi }{3}}}{\sqrt{3}} \\
& 3 & \nicefrac{1}{\sqrt{3}} & \nicefrac{e^{\nicefrac{-2 i \pi }{3}}}{\sqrt{3}} & \nicefrac{e^{\nicefrac{2 i \pi }{3}}}{\sqrt{3}} \\ \hline
\multirow{3}{*}{3} & 1 & \nicefrac{1}{\sqrt{3}} & \nicefrac{e^{\nicefrac{2 i \pi }{3}}}{\sqrt{3}} & \nicefrac{e^{\nicefrac{2 i \pi }{3}}}{\sqrt{3}} \\
& 2 & \nicefrac{1}{\sqrt{3}} & \nicefrac{e^{\nicefrac{-2 i \pi }{3}}}{\sqrt{3}} & \nicefrac{1}{\sqrt{3}} \\
& 3 & \nicefrac{1}{\sqrt{3}} & \nicefrac{1}{\sqrt{3}} & \nicefrac{e^{\nicefrac{-2 i \pi }{3}}}{\sqrt{3}} \\ \hline
\multirow{3}{*}{4} & 1 & \nicefrac{1}{\sqrt{3}} & \nicefrac{e^{\nicefrac{-2 i \pi }{3}}}{\sqrt{3}} & \nicefrac{e^{\nicefrac{-2 i \pi }{3}}}{\sqrt{3}} \\
& 2 & \nicefrac{1}{\sqrt{3}} & \nicefrac{1}{\sqrt{3}} & \nicefrac{e^{\nicefrac{2 i \pi }{3}}}{\sqrt{3}} \\
& 3 & \nicefrac{1}{\sqrt{3}} & \nicefrac{e^{\nicefrac{2 i \pi }{3}}}{\sqrt{3}} & \nicefrac{1}{\sqrt{3}}
\end{tabular}
\end{ruledtabular}
\vspace{1em}

\subsubsection{Coefficients for \(d=4\)}

\begin{ruledtabular}
\begin{tabular}{CCCCCC}
m & i & c_1 & c_2 & c_3 & c_4 \\ \hline
\multirow{4}{*}{1} & 1 & 1 & 0 & 0 & 0 \\
& 2 & 0 & 1 & 0 & 0 \\
& 3 & 0 & 0 & 1 & 0 \\
& 4 & 0 & 0 & 0 & 1 \\ \hline
\multirow{4}{*}{2} & 1 & \nicefrac{1}{2} & \nicefrac{1}{2} & \nicefrac{1}{2} & \nicefrac{1}{2} \\
& 2 & \nicefrac{1}{2} & \nicefrac{1}{2} & \nicefrac{-1}{2} & \nicefrac{-1}{2} \\
& 3 & \nicefrac{1}{2} & \nicefrac{-1}{2} & \nicefrac{-1}{2} & \nicefrac{1}{2} \\
& 4 & \nicefrac{1}{2} & \nicefrac{-1}{2} & \nicefrac{1}{2} & \nicefrac{-1}{2} \\ \hline
\multirow{4}{*}{3} & 1 & \nicefrac{1}{2} & \nicefrac{1}{2} & \nicefrac{-i}{2} & \nicefrac{i}{2} \\
& 2 & \nicefrac{1}{2} & \nicefrac{1}{2} & \nicefrac{i}{2} & \nicefrac{-i}{2} \\
& 3 & \nicefrac{1}{2} & \nicefrac{-1}{2} & \nicefrac{i}{2} & \nicefrac{i}{2} \\
& 4 & \nicefrac{1}{2} & \nicefrac{-1}{2} & \nicefrac{-i}{2} & \nicefrac{-i}{2} \\ \hline
\multirow{4}{*}{4} & 1 & \nicefrac{1}{2} & \nicefrac{i}{2} & \nicefrac{-1}{2} & \nicefrac{i}{2} \\
& 2 & \nicefrac{1}{2} & \nicefrac{-i}{2} & \nicefrac{-1}{2} & \nicefrac{-i}{2} \\
& 3 & \nicefrac{1}{2} & \nicefrac{i}{2} & \nicefrac{1}{2} & \nicefrac{-i}{2} \\
& 4 & \nicefrac{1}{2} & \nicefrac{-i}{2} & \nicefrac{1}{2} & \nicefrac{i}{2} \\ \hline
\multirow{4}{*}{5} & 1 & \nicefrac{1}{2} & \nicefrac{i}{2} & \nicefrac{i}{2} & \nicefrac{-1}{2} \\
& 2 & \nicefrac{1}{2} & \nicefrac{-i}{2} & \nicefrac{-i}{2} & \nicefrac{-1}{2} \\
& 3 & \nicefrac{1}{2} & \nicefrac{i}{2} & \nicefrac{-i}{2} & \nicefrac{1}{2} \\
& 4 & \nicefrac{1}{2} & \nicefrac{-i}{2} & \nicefrac{i}{2} & \nicefrac{1}{2}
\end{tabular}
\end{ruledtabular}
\vspace{1em}

\subsubsection{Coefficients for \(d=5\)}

\begin{ruledtabular}
\begin{tabular}{CCCCCCC}
m & i & c_1 & c_2 & c_3 & c_4 & c_5 \\ \hline
\multirow{5}{*}{1} & 1 & 1 & 0 & 0 & 0 & 0 \\
& 2 & 0 & 1 & 0 & 0 & 0 \\
& 3 & 0 & 0 & 1 & 0 & 0 \\
& 4 & 0 & 0 & 0 & 1 & 0 \\
& 5 & 0 & 0 & 0 & 0 & 1 \\ \hline
\multirow{5}{*}{2} & 1 & \nicefrac{1}{\sqrt{5}} & \nicefrac{1}{\sqrt{5}} & \nicefrac{1}{\sqrt{5}} & \nicefrac{1}{\sqrt{5}} & \nicefrac{1}{\sqrt{5}} \\
& 2 & \nicefrac{1}{\sqrt{5}} & \nicefrac{e^{\nicefrac{2 i \pi }{5}}}{\sqrt{5}} & \nicefrac{e^{\nicefrac{4 i \pi }{5}}}{\sqrt{5}} & \nicefrac{e^{\nicefrac{-4 i \pi }{5}}}{\sqrt{5}} & \nicefrac{e^{\nicefrac{-2 i \pi }{5}}}{\sqrt{5}} \\
& 3 & \nicefrac{1}{\sqrt{5}} & \nicefrac{e^{\nicefrac{4 i \pi }{5}}}{\sqrt{5}} & \nicefrac{e^{\nicefrac{-2 i \pi }{5}}}{\sqrt{5}} & \nicefrac{e^{\nicefrac{2 i \pi }{5}}}{\sqrt{5}} & \nicefrac{e^{\nicefrac{-4 i \pi }{5}}}{\sqrt{5}} \\
& 4 & \nicefrac{1}{\sqrt{5}} & \nicefrac{e^{\nicefrac{-4 i \pi }{5}}}{\sqrt{5}} & \nicefrac{e^{\nicefrac{2 i \pi }{5}}}{\sqrt{5}} & \nicefrac{e^{\nicefrac{-2 i \pi }{5}}}{\sqrt{5}} & \nicefrac{e^{\nicefrac{4 i \pi }{5}}}{\sqrt{5}} \\
& 5 & \nicefrac{1}{\sqrt{5}} & \nicefrac{e^{\nicefrac{-2 i \pi }{5}}}{\sqrt{5}} & \nicefrac{e^{\nicefrac{-4 i \pi }{5}}}{\sqrt{5}} & \nicefrac{e^{\nicefrac{4 i \pi }{5}}}{\sqrt{5}} & \nicefrac{e^{\nicefrac{2 i \pi }{5}}}{\sqrt{5}} \\ \hline
\multirow{5}{*}{3} & 1 & \nicefrac{1}{\sqrt{5}} & \nicefrac{e^{\nicefrac{2 i \pi }{5}}}{\sqrt{5}} & \nicefrac{e^{\nicefrac{-2 i \pi }{5}}}{\sqrt{5}} & \nicefrac{e^{\nicefrac{-2 i \pi }{5}}}{\sqrt{5}} & \nicefrac{e^{\nicefrac{2 i \pi }{5}}}{\sqrt{5}} \\
& 2 & \nicefrac{1}{\sqrt{5}} & \nicefrac{e^{\nicefrac{4 i \pi }{5}}}{\sqrt{5}} & \nicefrac{e^{\nicefrac{2 i \pi }{5}}}{\sqrt{5}} & \nicefrac{e^{\nicefrac{4 i \pi }{5}}}{\sqrt{5}} & \nicefrac{1}{\sqrt{5}} \\
& 3 & \nicefrac{1}{\sqrt{5}} & \nicefrac{e^{\nicefrac{-4 i \pi }{5}}}{\sqrt{5}} & \nicefrac{e^{\nicefrac{-4 i \pi }{5}}}{\sqrt{5}} & \nicefrac{1}{\sqrt{5}} & \nicefrac{e^{\nicefrac{-2 i \pi }{5}}}{\sqrt{5}} \\
& 4 & \nicefrac{1}{\sqrt{5}} & \nicefrac{e^{\nicefrac{-2 i \pi }{5}}}{\sqrt{5}} & \nicefrac{1}{\sqrt{5}} & \nicefrac{e^{\nicefrac{-4 i \pi }{5}}}{\sqrt{5}} & \nicefrac{e^{\nicefrac{-4 i \pi }{5}}}{\sqrt{5}} \\
& 5 & \nicefrac{1}{\sqrt{5}} & \nicefrac{1}{\sqrt{5}} & \nicefrac{e^{\nicefrac{4 i \pi }{5}}}{\sqrt{5}} & \nicefrac{e^{\nicefrac{2 i \pi }{5}}}{\sqrt{5}} & \nicefrac{e^{\nicefrac{4 i \pi }{5}}}{\sqrt{5}} \\ \hline
& 1 & \nicefrac{1}{\sqrt{5}} & \nicefrac{e^{\nicefrac{4 i \pi }{5}}}{\sqrt{5}} & \nicefrac{e^{\nicefrac{-4 i \pi }{5}}}{\sqrt{5}} & \nicefrac{e^{\nicefrac{-4 i \pi }{5}}}{\sqrt{5}} & \nicefrac{e^{\nicefrac{4 i \pi }{5}}}{\sqrt{5}} \\
\multirow{5}{*}{4} & 2 & \nicefrac{1}{\sqrt{5}} & \nicefrac{e^{\nicefrac{-4 i \pi }{5}}}{\sqrt{5}} & \nicefrac{1}{\sqrt{5}} & \nicefrac{e^{\nicefrac{2 i \pi }{5}}}{\sqrt{5}} & \nicefrac{e^{\nicefrac{2 i \pi }{5}}}{\sqrt{5}} \\
& 3 & \nicefrac{1}{\sqrt{5}} & \nicefrac{e^{\nicefrac{-2 i \pi }{5}}}{\sqrt{5}} & \nicefrac{e^{\nicefrac{4 i \pi }{5}}}{\sqrt{5}} & \nicefrac{e^{\nicefrac{-2 i \pi }{5}}}{\sqrt{5}} & \nicefrac{1}{\sqrt{5}} \\
& 4 & \nicefrac{1}{\sqrt{5}} & \nicefrac{1}{\sqrt{5}} & \nicefrac{e^{\nicefrac{-2 i \pi }{5}}}{\sqrt{5}} & \nicefrac{e^{\nicefrac{4 i \pi }{5}}}{\sqrt{5}} & \nicefrac{e^{\nicefrac{-2 i \pi }{5}}}{\sqrt{5}} \\
& 5 & \nicefrac{1}{\sqrt{5}} & \nicefrac{e^{\nicefrac{2 i \pi }{5}}}{\sqrt{5}} & \nicefrac{e^{\nicefrac{2 i \pi }{5}}}{\sqrt{5}} & \nicefrac{1}{\sqrt{5}} & \nicefrac{e^{\nicefrac{-4 i \pi }{5}}}{\sqrt{5}} \\ \hline
\multirow{5}{*}{5} & 1 & \nicefrac{1}{\sqrt{5}} & \nicefrac{e^{\nicefrac{-4 i \pi }{5}}}{\sqrt{5}} & \nicefrac{e^{\nicefrac{4 i \pi }{5}}}{\sqrt{5}} & \nicefrac{e^{\nicefrac{4 i \pi }{5}}}{\sqrt{5}} & \nicefrac{e^{\nicefrac{-4 i \pi }{5}}}{\sqrt{5}} \\
& 2 & \nicefrac{1}{\sqrt{5}} & \nicefrac{e^{\nicefrac{-2 i \pi }{5}}}{\sqrt{5}} & \nicefrac{e^{\nicefrac{-2 i \pi }{5}}}{\sqrt{5}} & \nicefrac{1}{\sqrt{5}} & \nicefrac{e^{\nicefrac{4 i \pi }{5}}}{\sqrt{5}} \\
& 3 & \nicefrac{1}{\sqrt{5}} & \nicefrac{1}{\sqrt{5}} & \nicefrac{e^{\nicefrac{2 i \pi }{5}}}{\sqrt{5}} & \nicefrac{e^{\nicefrac{-4 i \pi }{5}}}{\sqrt{5}} & \nicefrac{e^{\nicefrac{2 i \pi }{5}}}{\sqrt{5}} \\
& 4 & \nicefrac{1}{\sqrt{5}} & \nicefrac{e^{\nicefrac{2 i \pi }{5}}}{\sqrt{5}} & \nicefrac{e^{\nicefrac{-4 i \pi }{5}}}{\sqrt{5}} & \nicefrac{e^{\nicefrac{2 i \pi }{5}}}{\sqrt{5}} & \nicefrac{1}{\sqrt{5}} \\
& 5 & \nicefrac{1}{\sqrt{5}} & \nicefrac{e^{\nicefrac{4 i \pi }{5}}}{\sqrt{5}} & \nicefrac{1}{\sqrt{5}} & \nicefrac{e^{\nicefrac{-2 i \pi }{5}}}{\sqrt{5}} & \nicefrac{e^{\nicefrac{-2 i \pi }{5}}}{\sqrt{5}} \\ \hline
\multirow{5}{*}{6} & 1 & \nicefrac{1}{\sqrt{5}} & \nicefrac{e^{\nicefrac{-2 i \pi }{5}}}{\sqrt{5}} & \nicefrac{e^{\nicefrac{2 i \pi }{5}}}{\sqrt{5}} & \nicefrac{e^{\nicefrac{2 i \pi }{5}}}{\sqrt{5}} & \nicefrac{e^{\nicefrac{-2 i \pi }{5}}}{\sqrt{5}} \\
& 2 & \nicefrac{1}{\sqrt{5}} & \nicefrac{1}{\sqrt{5}} & \nicefrac{e^{\nicefrac{-4 i \pi }{5}}}{\sqrt{5}} & \nicefrac{e^{\nicefrac{-2 i \pi }{5}}}{\sqrt{5}} & \nicefrac{e^{\nicefrac{-4 i \pi }{5}}}{\sqrt{5}} \\
& 3 & \nicefrac{1}{\sqrt{5}} & \nicefrac{e^{\nicefrac{2 i \pi }{5}}}{\sqrt{5}} & \nicefrac{1}{\sqrt{5}} & \nicefrac{e^{\nicefrac{4 i \pi }{5}}}{\sqrt{5}} & \nicefrac{e^{\nicefrac{4 i \pi }{5}}}{\sqrt{5}} \\
& 4 & \nicefrac{1}{\sqrt{5}} & \nicefrac{e^{\nicefrac{4 i \pi }{5}}}{\sqrt{5}} & \nicefrac{e^{\nicefrac{4 i \pi }{5}}}{\sqrt{5}} & \nicefrac{1}{\sqrt{5}} & \nicefrac{e^{\nicefrac{2 i \pi }{5}}}{\sqrt{5}} \\
& 5 & \nicefrac{1}{\sqrt{5}} & \nicefrac{e^{\nicefrac{-4 i \pi }{5}}}{\sqrt{5}} & \nicefrac{e^{\nicefrac{-2 i \pi }{5}}}{\sqrt{5}} & \nicefrac{e^{\nicefrac{-4 i \pi }{5}}}{\sqrt{5}} & \nicefrac{1}{\sqrt{5}}
\end{tabular}
\end{ruledtabular}
\vspace{1em}

\bibliography{MUBs.bib}

\begin{thebibliography}{50}
\expandafter\ifx\csname natexlab\endcsname\relax\def\natexlab#1{#1}\fi
\expandafter\ifx\csname bibnamefont\endcsname\relax
  \def\bibnamefont#1{#1}\fi
\expandafter\ifx\csname bibfnamefont\endcsname\relax
  \def\bibfnamefont#1{#1}\fi
\expandafter\ifx\csname citenamefont\endcsname\relax
  \def\citenamefont#1{#1}\fi
\expandafter\ifx\csname url\endcsname\relax
  \def\url#1{\texttt{#1}}\fi
\expandafter\ifx\csname urlprefix\endcsname\relax\def\urlprefix{URL }\fi
\providecommand{\bibinfo}[2]{#2}
\providecommand{\eprint}[2][]{\url{#2}}

\bibitem[{\citenamefont{Wootters and Fields}(1989)}]{Wootters:1989}
\bibinfo{author}{\bibfnamefont{W.~K.} \bibnamefont{Wootters}} \bibnamefont{and}
  \bibinfo{author}{\bibfnamefont{B.~D.} \bibnamefont{Fields}},
  \bibinfo{journal}{Ann. Phys.} \textbf{\bibinfo{volume}{191}},
  \bibinfo{pages}{363} (\bibinfo{year}{1989}).

\bibitem[{\citenamefont{Ivanovi{\'c}}(1981)}]{Ivanovic:1981}
\bibinfo{author}{\bibfnamefont{I.~D.} \bibnamefont{Ivanovi{\'c}}},
  \bibinfo{journal}{J. Phys. A} \textbf{\bibinfo{volume}{14}},
  \bibinfo{pages}{3241} (\bibinfo{year}{1981}).

\bibitem[{\citenamefont{Wehner and Winter}(2010)}]{Wehner:2010}
\bibinfo{author}{\bibfnamefont{S.}~\bibnamefont{Wehner}} \bibnamefont{and}
  \bibinfo{author}{\bibfnamefont{A.}~\bibnamefont{Winter}},
  \bibinfo{journal}{New J. Phys.} \textbf{\bibinfo{volume}{12}},
  \bibinfo{pages}{025009} (\bibinfo{year}{2010}).

\bibitem[{\citenamefont{Barnett}(2009)}]{Barnett:2009}
\bibinfo{author}{\bibfnamefont{S.~M.} \bibnamefont{Barnett}},
  \emph{\bibinfo{title}{Quantum Information}} (\bibinfo{publisher}{Oxford
  University Press}, \bibinfo{year}{2009}).

\bibitem[{\citenamefont{Durt et~al.}(2010)\citenamefont{Durt, Englert,
  Bengtsson, and {\.Z}yczkowski}}]{Durt:2010a}
\bibinfo{author}{\bibfnamefont{T.}~\bibnamefont{Durt}},
  \bibinfo{author}{\bibfnamefont{B.-G.} \bibnamefont{Englert}},
  \bibinfo{author}{\bibfnamefont{I.}~\bibnamefont{Bengtsson}},
  \bibnamefont{and}
  \bibinfo{author}{\bibfnamefont{K.}~\bibnamefont{{\.Z}yczkowski}},
  \bibinfo{journal}{International Journal of Quantum Information}
  \textbf{\bibinfo{volume}{08}}, \bibinfo{pages}{535} (\bibinfo{year}{2010}).

\bibitem[{\citenamefont{Bennett and Brassard}(1984)}]{Bennett:1984}
\bibinfo{author}{\bibfnamefont{C.~H.} \bibnamefont{Bennett}} \bibnamefont{and}
  \bibinfo{author}{\bibfnamefont{G.}~\bibnamefont{Brassard}}, in
  \emph{\bibinfo{booktitle}{Proceedings of the IEEE International Conference on
  Computers, Systems, and Signal Processing, Bangalore}}
  (\bibinfo{year}{1984}), p. \bibinfo{pages}{175}.

\bibitem[{\citenamefont{Jun-Lin and Chuan}(2010)}]{Jun-Lin:2010}
\bibinfo{author}{\bibfnamefont{L.}~\bibnamefont{Jun-Lin}} \bibnamefont{and}
  \bibinfo{author}{\bibfnamefont{W.}~\bibnamefont{Chuan}},
  \bibinfo{journal}{Chinese Physics Letters} \textbf{\bibinfo{volume}{27}},
  \bibinfo{pages}{110303} (\bibinfo{year}{2010}).

\bibitem[{\citenamefont{Malik et~al.}(2012)\citenamefont{Malik, O'Sullivan,
  Rodenburg, Mirhosseini, Leach, Lavery, Padgett, and Boyd}}]{Malik:2012a}
\bibinfo{author}{\bibfnamefont{M.}~\bibnamefont{Malik}},
  \bibinfo{author}{\bibfnamefont{M.}~\bibnamefont{O'Sullivan}},
  \bibinfo{author}{\bibfnamefont{B.}~\bibnamefont{Rodenburg}},
  \bibinfo{author}{\bibfnamefont{M.}~\bibnamefont{Mirhosseini}},
  \bibinfo{author}{\bibfnamefont{J.}~\bibnamefont{Leach}},
  \bibinfo{author}{\bibfnamefont{M.~P.~J.} \bibnamefont{Lavery}},
  \bibinfo{author}{\bibfnamefont{M.~J.} \bibnamefont{Padgett}},
  \bibnamefont{and} \bibinfo{author}{\bibfnamefont{R.~W.} \bibnamefont{Boyd}},
  \bibinfo{journal}{Opt. Express} \textbf{\bibinfo{volume}{20}},
  \bibinfo{pages}{13195} (\bibinfo{year}{2012}).

\bibitem[{\citenamefont{Filippov and Man'ko}(2011)}]{Filippov:2011}
\bibinfo{author}{\bibfnamefont{S.~N.} \bibnamefont{Filippov}} \bibnamefont{and}
  \bibinfo{author}{\bibfnamefont{V.~I.} \bibnamefont{Man'ko}},
  \bibinfo{journal}{Physica Scripta} \textbf{\bibinfo{volume}{2011}},
  \bibinfo{pages}{014010} (\bibinfo{year}{2011}).

\bibitem[{\citenamefont{Fern\'andez-P\'erez
  et~al.}(2011)\citenamefont{Fern\'andez-P\'erez, Klimov, and
  Saavedra}}]{Fernandez-Perez:2011}
\bibinfo{author}{\bibfnamefont{A.}~\bibnamefont{Fern\'andez-P\'erez}},
  \bibinfo{author}{\bibfnamefont{A.~B.} \bibnamefont{Klimov}},
  \bibnamefont{and} \bibinfo{author}{\bibfnamefont{C.}~\bibnamefont{Saavedra}},
  \bibinfo{journal}{Phys. Rev. A} \textbf{\bibinfo{volume}{83}},
  \bibinfo{pages}{052332} (\bibinfo{year}{2011}).

\bibitem[{\citenamefont{Adamson and Steinberg}(2010)}]{Adamson:2010}
\bibinfo{author}{\bibfnamefont{R.~B.~A.} \bibnamefont{Adamson}}
  \bibnamefont{and} \bibinfo{author}{\bibfnamefont{A.~M.}
  \bibnamefont{Steinberg}}, \bibinfo{journal}{Phys. Rev. Lett.}
  \textbf{\bibinfo{volume}{105}}, \bibinfo{pages}{030406}
  (\bibinfo{year}{2010}).

\bibitem[{\citenamefont{Lima et~al.}(2011)\citenamefont{Lima, Neves,
  Guzm\'{a}n, G\'{o}mez, Nogueira, Delgado, Vargas, and Saavedra}}]{Lima:2011}
\bibinfo{author}{\bibfnamefont{G.}~\bibnamefont{Lima}},
  \bibinfo{author}{\bibfnamefont{L.}~\bibnamefont{Neves}},
  \bibinfo{author}{\bibfnamefont{R.}~\bibnamefont{Guzm\'{a}n}},
  \bibinfo{author}{\bibfnamefont{E.~S.} \bibnamefont{G\'{o}mez}},
  \bibinfo{author}{\bibfnamefont{W.~A.~T.} \bibnamefont{Nogueira}},
  \bibinfo{author}{\bibfnamefont{A.}~\bibnamefont{Delgado}},
  \bibinfo{author}{\bibfnamefont{A.}~\bibnamefont{Vargas}}, \bibnamefont{and}
  \bibinfo{author}{\bibfnamefont{C.}~\bibnamefont{Saavedra}},
  \bibinfo{journal}{Opt. Express} \textbf{\bibinfo{volume}{19}},
  \bibinfo{pages}{3542} (\bibinfo{year}{2011}).

\bibitem[{\citenamefont{Bandyopadhyay et~al.}(2002)\citenamefont{Bandyopadhyay,
  Boykin, Roychowdhury, and Vatan}}]{Bandyopadhyay:2002}
\bibinfo{author}{\bibfnamefont{S.}~\bibnamefont{Bandyopadhyay}},
  \bibinfo{author}{\bibfnamefont{P.~O.} \bibnamefont{Boykin}},
  \bibinfo{author}{\bibfnamefont{V.}~\bibnamefont{Roychowdhury}},
  \bibnamefont{and} \bibinfo{author}{\bibfnamefont{F.}~\bibnamefont{Vatan}},
  \bibinfo{journal}{Algorithmica} \textbf{\bibinfo{volume}{34}},
  \bibinfo{pages}{512} (\bibinfo{year}{2002}).

\bibitem[{\citenamefont{Brierley and Weigert}(2010)}]{Brierley:2010a}
\bibinfo{author}{\bibfnamefont{S.}~\bibnamefont{Brierley}} \bibnamefont{and}
  \bibinfo{author}{\bibfnamefont{S.}~\bibnamefont{Weigert}},
  \bibinfo{journal}{Journal of Physics: Conference Series}
  \textbf{\bibinfo{volume}{254}}, \bibinfo{pages}{012008}
  (\bibinfo{year}{2010}).

\bibitem[{\citenamefont{Nagali et~al.}(2010)\citenamefont{Nagali, Sansoni,
  Marrucci, Santamato, and Sciarrino}}]{Nagali:2010a}
\bibinfo{author}{\bibfnamefont{E.}~\bibnamefont{Nagali}},
  \bibinfo{author}{\bibfnamefont{L.}~\bibnamefont{Sansoni}},
  \bibinfo{author}{\bibfnamefont{L.}~\bibnamefont{Marrucci}},
  \bibinfo{author}{\bibfnamefont{E.}~\bibnamefont{Santamato}},
  \bibnamefont{and}
  \bibinfo{author}{\bibfnamefont{F.}~\bibnamefont{Sciarrino}},
  \bibinfo{journal}{Phys. Rev. A} \textbf{\bibinfo{volume}{81}},
  \bibinfo{pages}{052317} (\bibinfo{year}{2010}).

\bibitem[{\citenamefont{Thew et~al.}(2002)\citenamefont{Thew, Nemoto, White,
  and Munro}}]{Thew:2002a}
\bibinfo{author}{\bibfnamefont{R.~T.} \bibnamefont{Thew}},
  \bibinfo{author}{\bibfnamefont{K.}~\bibnamefont{Nemoto}},
  \bibinfo{author}{\bibfnamefont{A.~G.} \bibnamefont{White}}, \bibnamefont{and}
  \bibinfo{author}{\bibfnamefont{W.~J.} \bibnamefont{Munro}},
  \bibinfo{journal}{Phys. Rev. A} \textbf{\bibinfo{volume}{66}},
  \bibinfo{pages}{012303} (\bibinfo{year}{2002}).

\bibitem[{\citenamefont{Yao and Padgett}(2011)}]{Yao:2011a}
\bibinfo{author}{\bibfnamefont{A.~M.} \bibnamefont{Yao}} \bibnamefont{and}
  \bibinfo{author}{\bibfnamefont{M.~J.} \bibnamefont{Padgett}},
  \bibinfo{journal}{Adv. Opt. Photon.} \textbf{\bibinfo{volume}{3}},
  \bibinfo{pages}{161} (\bibinfo{year}{2011}).

\bibitem[{\citenamefont{Allen et~al.}(1992)\citenamefont{Allen, Beijersbergen,
  Spreeuw, and Woerdman}}]{Allen:1992}
\bibinfo{author}{\bibfnamefont{L.}~\bibnamefont{Allen}},
  \bibinfo{author}{\bibfnamefont{M.~W.} \bibnamefont{Beijersbergen}},
  \bibinfo{author}{\bibfnamefont{R.~J.~C.} \bibnamefont{Spreeuw}},
  \bibnamefont{and} \bibinfo{author}{\bibfnamefont{J.~P.}
  \bibnamefont{Woerdman}}, \bibinfo{journal}{Phys. Rev. A}
  \textbf{\bibinfo{volume}{95}}, \bibinfo{pages}{8185} (\bibinfo{year}{1992}).

\bibitem[{\citenamefont{Mair et~al.}(2001)\citenamefont{Mair, Alipasha, Weihs,
  and Zeilinger}}]{Mair:2001}
\bibinfo{author}{\bibfnamefont{A.}~\bibnamefont{Mair}},
  \bibinfo{author}{\bibfnamefont{V.}~\bibnamefont{Alipasha}},
  \bibinfo{author}{\bibfnamefont{G.}~\bibnamefont{Weihs}}, \bibnamefont{and}
  \bibinfo{author}{\bibfnamefont{A.}~\bibnamefont{Zeilinger}},
  \bibinfo{journal}{Nature} \textbf{\bibinfo{volume}{412}},
  \bibinfo{pages}{313} (\bibinfo{year}{2001}).

\bibitem[{\citenamefont{Leach et~al.}(2010)\citenamefont{Leach, Jack, Romero,
  Jha, Yao, Franke-Arnold, Ireland, Boyd, Barnett, and Padgett}}]{Leach:2010}
\bibinfo{author}{\bibfnamefont{J.}~\bibnamefont{Leach}},
  \bibinfo{author}{\bibfnamefont{B.}~\bibnamefont{Jack}},
  \bibinfo{author}{\bibfnamefont{J.}~\bibnamefont{Romero}},
  \bibinfo{author}{\bibfnamefont{A.~K.} \bibnamefont{Jha}},
  \bibinfo{author}{\bibfnamefont{A.~M.} \bibnamefont{Yao}},
  \bibinfo{author}{\bibfnamefont{S.}~\bibnamefont{Franke-Arnold}},
  \bibinfo{author}{\bibfnamefont{D.~G.} \bibnamefont{Ireland}},
  \bibinfo{author}{\bibfnamefont{R.~W.} \bibnamefont{Boyd}},
  \bibinfo{author}{\bibfnamefont{S.~M.} \bibnamefont{Barnett}},
  \bibnamefont{and} \bibinfo{author}{\bibfnamefont{M.~J.}
  \bibnamefont{Padgett}}, \bibinfo{journal}{Science}
  \textbf{\bibinfo{volume}{329}}, \bibinfo{pages}{662} (\bibinfo{year}{2010}).

\bibitem[{\citenamefont{Gr{\"o}blacher
  et~al.}(2006)\citenamefont{Gr{\"o}blacher, Jennewein, Vaziri, Weihs, and
  Zeilinger}}]{Groblacher:2006}
\bibinfo{author}{\bibfnamefont{S.}~\bibnamefont{Gr{\"o}blacher}},
  \bibinfo{author}{\bibfnamefont{T.}~\bibnamefont{Jennewein}},
  \bibinfo{author}{\bibfnamefont{A.}~\bibnamefont{Vaziri}},
  \bibinfo{author}{\bibfnamefont{G.}~\bibnamefont{Weihs}}, \bibnamefont{and}
  \bibinfo{author}{\bibfnamefont{A.}~\bibnamefont{Zeilinger}},
  \bibinfo{journal}{New J. Phys.} \textbf{\bibinfo{volume}{8}},
  \bibinfo{pages}{75} (\bibinfo{year}{2006}).

\bibitem[{\citenamefont{Salakhutdinov et~al.}(2012)\citenamefont{Salakhutdinov,
  Eliel, and L\"offler}}]{Salakhutdinov:2012}
\bibinfo{author}{\bibfnamefont{V.~D.} \bibnamefont{Salakhutdinov}},
  \bibinfo{author}{\bibfnamefont{E.~R.} \bibnamefont{Eliel}}, \bibnamefont{and}
  \bibinfo{author}{\bibfnamefont{W.}~\bibnamefont{L\"offler}},
  \bibinfo{journal}{Phys. Rev. Lett.} \textbf{\bibinfo{volume}{108}},
  \bibinfo{pages}{173604} (\bibinfo{year}{2012}).

\bibitem[{\citenamefont{Bourennane et~al.}(2002)\citenamefont{Bourennane,
  Karlsson, Bj{\"o}rk, Gisin, and Cerf}}]{Bourennane:2002}
\bibinfo{author}{\bibfnamefont{M.}~\bibnamefont{Bourennane}},
  \bibinfo{author}{\bibfnamefont{A.}~\bibnamefont{Karlsson}},
  \bibinfo{author}{\bibfnamefont{G.}~\bibnamefont{Bj{\"o}rk}},
  \bibinfo{author}{\bibfnamefont{N.}~\bibnamefont{Gisin}}, \bibnamefont{and}
  \bibinfo{author}{\bibfnamefont{N.~J.} \bibnamefont{Cerf}},
  \bibinfo{journal}{Journal of Physics A: Mathematical and General}
  \textbf{\bibinfo{volume}{35}}, \bibinfo{pages}{10065} (\bibinfo{year}{2002}).

\bibitem[{\citenamefont{Cerf et~al.}(2002)\citenamefont{Cerf, Bourennane,
  Karlsson, and Gisin}}]{Cerf:2002}
\bibinfo{author}{\bibfnamefont{N.~J.} \bibnamefont{Cerf}},
  \bibinfo{author}{\bibfnamefont{M.}~\bibnamefont{Bourennane}},
  \bibinfo{author}{\bibfnamefont{A.}~\bibnamefont{Karlsson}}, \bibnamefont{and}
  \bibinfo{author}{\bibfnamefont{N.}~\bibnamefont{Gisin}},
  \bibinfo{journal}{Phys. Rev. Lett.} \textbf{\bibinfo{volume}{88}},
  \bibinfo{pages}{127902} (\bibinfo{year}{2002}).

\bibitem[{\citenamefont{Bechmann-Pasquinucci and
  Tittel}(2000)}]{Bechmann-Pasquinucci:2000}
\bibinfo{author}{\bibfnamefont{H.}~\bibnamefont{Bechmann-Pasquinucci}}
  \bibnamefont{and} \bibinfo{author}{\bibfnamefont{W.}~\bibnamefont{Tittel}},
  \bibinfo{journal}{Phys. Rev. A} \textbf{\bibinfo{volume}{61}},
  \bibinfo{pages}{062308} (\bibinfo{year}{2000}).

\bibitem[{\citenamefont{Walborn et~al.}(2006)\citenamefont{Walborn, Lemelle,
  Almeida, and Ribeiro}}]{Walborn:2006}
\bibinfo{author}{\bibfnamefont{S.~P.} \bibnamefont{Walborn}},
  \bibinfo{author}{\bibfnamefont{D.~S.} \bibnamefont{Lemelle}},
  \bibinfo{author}{\bibfnamefont{M.~P.} \bibnamefont{Almeida}},
  \bibnamefont{and} \bibinfo{author}{\bibfnamefont{P.~H.~S.}
  \bibnamefont{Ribeiro}}, \bibinfo{journal}{Phys. Rev. Lett.}
  \textbf{\bibinfo{volume}{96}}, \bibinfo{pages}{090501}
  (\bibinfo{year}{2006}).

\bibitem[{\citenamefont{Dixon et~al.}(2012)\citenamefont{Dixon, Howland,
  Schneeloch, and Howell}}]{Dixon:2012}
\bibinfo{author}{\bibfnamefont{P.~B.} \bibnamefont{Dixon}},
  \bibinfo{author}{\bibfnamefont{G.~A.} \bibnamefont{Howland}},
  \bibinfo{author}{\bibfnamefont{J.}~\bibnamefont{Schneeloch}},
  \bibnamefont{and} \bibinfo{author}{\bibfnamefont{J.~C.}
  \bibnamefont{Howell}}, \bibinfo{journal}{Phys. Rev. Lett.}
  \textbf{\bibinfo{volume}{108}}, \bibinfo{pages}{143603}
  (\bibinfo{year}{2012}).

\bibitem[{\citenamefont{Gruneisen et~al.}(2012)\citenamefont{Gruneisen, Black,
  Dymale, and Stoltenberg}}]{Gruneisen:2012}
\bibinfo{author}{\bibfnamefont{M.~T.} \bibnamefont{Gruneisen}},
  \bibinfo{author}{\bibfnamefont{J.~P.} \bibnamefont{Black}},
  \bibinfo{author}{\bibfnamefont{R.~C.} \bibnamefont{Dymale}},
  \bibnamefont{and} \bibinfo{author}{\bibfnamefont{K.~E.}
  \bibnamefont{Stoltenberg}}, in \emph{\bibinfo{booktitle}{Proc. SPIE 8542,
  Electro-Optical Remote Sensing, Photonic Technologies, and Applications VI,
  85421Q}} (\bibinfo{year}{2012}), pp. \bibinfo{pages}{85421Q--85421Q--14}.

\bibitem[{\citenamefont{Dada et~al.}(2011)\citenamefont{Dada, Leach, Buller,
  Padgett, and Andersson}}]{Dada:2011}
\bibinfo{author}{\bibfnamefont{A.~C.} \bibnamefont{Dada}},
  \bibinfo{author}{\bibfnamefont{J.}~\bibnamefont{Leach}},
  \bibinfo{author}{\bibfnamefont{G.~S.} \bibnamefont{Buller}},
  \bibinfo{author}{\bibfnamefont{M.~J.} \bibnamefont{Padgett}},
  \bibnamefont{and}
  \bibinfo{author}{\bibfnamefont{E.}~\bibnamefont{Andersson}},
  \bibinfo{journal}{Nature Physics} \textbf{\bibinfo{volume}{7}},
  \bibinfo{pages}{677} (\bibinfo{year}{2011}).

\bibitem[{\citenamefont{Gruneisen et~al.}(2008)\citenamefont{Gruneisen, Miller,
  Dymale, and Sweiti}}]{Gruneisen:2008}
\bibinfo{author}{\bibfnamefont{M.~T.} \bibnamefont{Gruneisen}},
  \bibinfo{author}{\bibfnamefont{W.~A.} \bibnamefont{Miller}},
  \bibinfo{author}{\bibfnamefont{R.~C.} \bibnamefont{Dymale}},
  \bibnamefont{and} \bibinfo{author}{\bibfnamefont{A.~M.}
  \bibnamefont{Sweiti}}, \bibinfo{journal}{Appl. Opt.}
  \textbf{\bibinfo{volume}{47}}, \bibinfo{pages}{A32} (\bibinfo{year}{2008}).

\bibitem[{\citenamefont{Schwinger}(1960)}]{Schwinger:1960}
\bibinfo{author}{\bibfnamefont{J.}~\bibnamefont{Schwinger}},
  \bibinfo{journal}{Proc. Natl. Acad. Sci. USA} \textbf{\bibinfo{volume}{46}},
  \bibinfo{pages}{570} (\bibinfo{year}{1960}).

\bibitem[{\citenamefont{Klappenecker and
  R{\"o}tteler}(2004)}]{Klappenecker:2004}
\bibinfo{author}{\bibfnamefont{A.}~\bibnamefont{Klappenecker}}
  \bibnamefont{and}
  \bibinfo{author}{\bibfnamefont{M.}~\bibnamefont{R{\"o}tteler}}, in
  \emph{\bibinfo{booktitle}{Finite Fields and Applications}}
  (\bibinfo{publisher}{Springer}, \bibinfo{year}{2004}), vol.
  \bibinfo{volume}{2948/2004}, pp. \bibinfo{pages}{262--266}.

\bibitem[{\citenamefont{Brierley et~al.}(2010)\citenamefont{Brierley, Weigert,
  and Bengtsson}}]{Brierley:2010}
\bibinfo{author}{\bibfnamefont{S.}~\bibnamefont{Brierley}},
  \bibinfo{author}{\bibfnamefont{S.}~\bibnamefont{Weigert}}, \bibnamefont{and}
  \bibinfo{author}{\bibfnamefont{I.}~\bibnamefont{Bengtsson}},
  \bibinfo{journal}{Quantum Inf. Comput.} \textbf{\bibinfo{volume}{10}},
  \bibinfo{pages}{0803} (\bibinfo{year}{2010}).

\bibitem[{\citenamefont{Klimov et~al.}(2008)\citenamefont{Klimov, Mu{\~n}oz,
  Fern{\'a}ndez, and Saavedra}}]{Klimov:2008}
\bibinfo{author}{\bibfnamefont{A.~B.} \bibnamefont{Klimov}},
  \bibinfo{author}{\bibfnamefont{C.}~\bibnamefont{Mu{\~n}oz}},
  \bibinfo{author}{\bibfnamefont{A.}~\bibnamefont{Fern{\'a}ndez}},
  \bibnamefont{and} \bibinfo{author}{\bibfnamefont{C.}~\bibnamefont{Saavedra}},
  \bibinfo{journal}{Phys. Rev. A} \textbf{\bibinfo{volume}{77}},
  \bibinfo{pages}{060303} (\bibinfo{year}{2008}).

\bibitem[{\citenamefont{James et~al.}(2001)\citenamefont{James, Kwiat, Munro,
  and White}}]{James:2001}
\bibinfo{author}{\bibfnamefont{D.~F.~V.} \bibnamefont{James}},
  \bibinfo{author}{\bibfnamefont{P.~G.} \bibnamefont{Kwiat}},
  \bibinfo{author}{\bibfnamefont{W.~J.} \bibnamefont{Munro}}, \bibnamefont{and}
  \bibinfo{author}{\bibfnamefont{A.~G.} \bibnamefont{White}},
  \bibinfo{journal}{Phys. Rev. A} \textbf{\bibinfo{volume}{64}},
  \bibinfo{pages}{052312} (\bibinfo{year}{2001}).

\bibitem[{\citenamefont{Langford et~al.}(2004)\citenamefont{Langford, Dalton,
  Harvey, O'Brien, Pryde, Gilchrist, Bartlett, and White}}]{Langford:2004}
\bibinfo{author}{\bibfnamefont{N.~K.} \bibnamefont{Langford}},
  \bibinfo{author}{\bibfnamefont{R.~B.} \bibnamefont{Dalton}},
  \bibinfo{author}{\bibfnamefont{M.~D.} \bibnamefont{Harvey}},
  \bibinfo{author}{\bibfnamefont{J.~L.} \bibnamefont{O'Brien}},
  \bibinfo{author}{\bibfnamefont{G.~J.} \bibnamefont{Pryde}},
  \bibinfo{author}{\bibfnamefont{A.}~\bibnamefont{Gilchrist}},
  \bibinfo{author}{\bibfnamefont{S.~D.} \bibnamefont{Bartlett}},
  \bibnamefont{and} \bibinfo{author}{\bibfnamefont{A.~G.} \bibnamefont{White}},
  \bibinfo{journal}{Phys. Rev. Lett.} \textbf{\bibinfo{volume}{93}},
  \bibinfo{pages}{053601} (\bibinfo{year}{2004}).

\bibitem[{\citenamefont{Altepeter et~al.}(2005)\citenamefont{Altepeter,
  Jeffrey, and Kwiat}}]{Altepeter:2005}
\bibinfo{author}{\bibfnamefont{J.~B.} \bibnamefont{Altepeter}},
  \bibinfo{author}{\bibfnamefont{E.~R.} \bibnamefont{Jeffrey}},
  \bibnamefont{and} \bibinfo{author}{\bibfnamefont{P.~G.} \bibnamefont{Kwiat}},
  in \emph{\bibinfo{booktitle}{Advances in Atomic, Molecular, and Optical
  Physics}} (\bibinfo{publisher}{Elsevier}, \bibinfo{year}{2005}).

\bibitem[{Sup()}]{SupplementalMaterial}
\bibinfo{note}{See Supplemental Material for further details on the
  normalization of the experimental counts, the completeness of the tomographic
  method, the numerical reconstruction procedure and a list of the MUBs
  employed.}

\bibitem[{\citenamefont{Romero et~al.}(2012)\citenamefont{Romero, Giovannini,
  Franke-Arnold, Barnett, and Padgett}}]{Romero:2012b}
\bibinfo{author}{\bibfnamefont{J.}~\bibnamefont{Romero}},
  \bibinfo{author}{\bibfnamefont{D.}~\bibnamefont{Giovannini}},
  \bibinfo{author}{\bibfnamefont{S.}~\bibnamefont{Franke-Arnold}},
  \bibinfo{author}{\bibfnamefont{S.~M.} \bibnamefont{Barnett}},
  \bibnamefont{and} \bibinfo{author}{\bibfnamefont{M.~J.}
  \bibnamefont{Padgett}}, \bibinfo{journal}{Phys. Rev. A}
  \textbf{\bibinfo{volume}{86}}, \bibinfo{pages}{012334}
  (\bibinfo{year}{2012}).

\bibitem[{\citenamefont{Arriz\'{o}n et~al.}(2007)\citenamefont{Arriz\'{o}n,
  Ruiz, Carrada, and Gonz\'{a}lez}}]{Arrizon:2007}
\bibinfo{author}{\bibfnamefont{V.}~\bibnamefont{Arriz\'{o}n}},
  \bibinfo{author}{\bibfnamefont{U.}~\bibnamefont{Ruiz}},
  \bibinfo{author}{\bibfnamefont{R.}~\bibnamefont{Carrada}}, \bibnamefont{and}
  \bibinfo{author}{\bibfnamefont{L.~A.} \bibnamefont{Gonz\'{a}lez}},
  \bibinfo{journal}{J. Opt. Soc. Am. A} \textbf{\bibinfo{volume}{24}},
  \bibinfo{pages}{3500} (\bibinfo{year}{2007}).

\bibitem[{\citenamefont{Davis et~al.}(1999)\citenamefont{Davis, Cottrell,
  Campos, Yzuel, and Moreno}}]{Davis:1999}
\bibinfo{author}{\bibfnamefont{J.~A.} \bibnamefont{Davis}},
  \bibinfo{author}{\bibfnamefont{D.~M.} \bibnamefont{Cottrell}},
  \bibinfo{author}{\bibfnamefont{J.}~\bibnamefont{Campos}},
  \bibinfo{author}{\bibfnamefont{M.~J.} \bibnamefont{Yzuel}}, \bibnamefont{and}
  \bibinfo{author}{\bibfnamefont{I.}~\bibnamefont{Moreno}},
  \bibinfo{journal}{Appl. Opt.} \textbf{\bibinfo{volume}{38}},
  \bibinfo{pages}{5004} (\bibinfo{year}{1999}).

\bibitem[{\citenamefont{Brierley and Weigert}(2009)}]{Brierley:2009a}
\bibinfo{author}{\bibfnamefont{S.}~\bibnamefont{Brierley}} \bibnamefont{and}
  \bibinfo{author}{\bibfnamefont{S.}~\bibnamefont{Weigert}},
  \bibinfo{journal}{Phys. Rev. A} \textbf{\bibinfo{volume}{79}},
  \bibinfo{pages}{052316} (\bibinfo{year}{2009}).

\bibitem[{\citenamefont{Padgett and Courtial}(1999)}]{Padgett:1999}
\bibinfo{author}{\bibfnamefont{M.~J.} \bibnamefont{Padgett}} \bibnamefont{and}
  \bibinfo{author}{\bibfnamefont{J.}~\bibnamefont{Courtial}},
  \bibinfo{journal}{Opt. Lett.} \textbf{\bibinfo{volume}{24}},
  \bibinfo{pages}{430} (\bibinfo{year}{1999}).

\bibitem[{\citenamefont{Bru\ss{}}(1998)}]{Bruss:1998}
\bibinfo{author}{\bibfnamefont{D.}~\bibnamefont{Bru\ss{}}},
  \bibinfo{journal}{Phys. Rev. Lett.} \textbf{\bibinfo{volume}{81}},
  \bibinfo{pages}{3018} (\bibinfo{year}{1998}).

\bibitem[{\citenamefont{Opatrn\'y et~al.}(1997)\citenamefont{Opatrn\'y, Welsch,
  and Vogel}}]{Opatrny:1997}
\bibinfo{author}{\bibfnamefont{T.}~\bibnamefont{Opatrn\'y}},
  \bibinfo{author}{\bibfnamefont{D.-G.} \bibnamefont{Welsch}},
  \bibnamefont{and} \bibinfo{author}{\bibfnamefont{W.}~\bibnamefont{Vogel}},
  \bibinfo{journal}{Phys. Rev. A} \textbf{\bibinfo{volume}{56}},
  \bibinfo{pages}{1788} (\bibinfo{year}{1997}).

\bibitem[{\citenamefont{Banaszek et~al.}(1999)\citenamefont{Banaszek, D'Ariano,
  Paris, and Sacchi}}]{Banaszek:1999}
\bibinfo{author}{\bibfnamefont{K.}~\bibnamefont{Banaszek}},
  \bibinfo{author}{\bibfnamefont{G.~M.} \bibnamefont{D'Ariano}},
  \bibinfo{author}{\bibfnamefont{M.~G.~A.} \bibnamefont{Paris}},
  \bibnamefont{and} \bibinfo{author}{\bibfnamefont{M.~F.}
  \bibnamefont{Sacchi}}, \bibinfo{journal}{Phys. Rev. A}
  \textbf{\bibinfo{volume}{61}}, \bibinfo{pages}{010304}
  (\bibinfo{year}{1999}).

\bibitem[{\citenamefont{Torres et~al.}(2003)\citenamefont{Torres, Alexandrescu,
  and Torner}}]{Torres:2003a}
\bibinfo{author}{\bibfnamefont{J.~P.} \bibnamefont{Torres}},
  \bibinfo{author}{\bibfnamefont{A.}~\bibnamefont{Alexandrescu}},
  \bibnamefont{and} \bibinfo{author}{\bibfnamefont{L.}~\bibnamefont{Torner}},
  \bibinfo{journal}{Phys. Rev. A} \textbf{\bibinfo{volume}{68}},
  \bibinfo{pages}{050301} (\bibinfo{year}{2003}).

\bibitem[{\citenamefont{Agnew et~al.}(2011)\citenamefont{Agnew, Leach, McLaren,
  Roux, and Boyd}}]{Agnew:2011}
\bibinfo{author}{\bibfnamefont{M.}~\bibnamefont{Agnew}},
  \bibinfo{author}{\bibfnamefont{J.}~\bibnamefont{Leach}},
  \bibinfo{author}{\bibfnamefont{M.}~\bibnamefont{McLaren}},
  \bibinfo{author}{\bibfnamefont{F.~S.} \bibnamefont{Roux}}, \bibnamefont{and}
  \bibinfo{author}{\bibfnamefont{R.~W.} \bibnamefont{Boyd}},
  \bibinfo{journal}{Phys. Rev. A} \textbf{\bibinfo{volume}{84}},
  \bibinfo{pages}{062101} (\bibinfo{year}{2011}).

\bibitem[{\citenamefont{Gross et~al.}(2010)\citenamefont{Gross, Liu, Flammia,
  Becker, and Eisert}}]{Gross:2010}
\bibinfo{author}{\bibfnamefont{D.}~\bibnamefont{Gross}},
  \bibinfo{author}{\bibfnamefont{Y.-K.} \bibnamefont{Liu}},
  \bibinfo{author}{\bibfnamefont{S.~T.} \bibnamefont{Flammia}},
  \bibinfo{author}{\bibfnamefont{S.}~\bibnamefont{Becker}}, \bibnamefont{and}
  \bibinfo{author}{\bibfnamefont{J.}~\bibnamefont{Eisert}},
  \bibinfo{journal}{Phys. Rev. Lett.} \textbf{\bibinfo{volume}{105}},
  \bibinfo{pages}{150401} (\bibinfo{year}{2010}).

\bibitem[{\citenamefont{Liu et~al.}(2012)\citenamefont{Liu, Zhang, Liu, Chen,
  and Yuan}}]{Liu:2012}
\bibinfo{author}{\bibfnamefont{W.-T.} \bibnamefont{Liu}},
  \bibinfo{author}{\bibfnamefont{T.}~\bibnamefont{Zhang}},
  \bibinfo{author}{\bibfnamefont{J.-Y.} \bibnamefont{Liu}},
  \bibinfo{author}{\bibfnamefont{P.-X.} \bibnamefont{Chen}}, \bibnamefont{and}
  \bibinfo{author}{\bibfnamefont{J.-M.} \bibnamefont{Yuan}},
  \bibinfo{journal}{Phys. Rev. Lett.} \textbf{\bibinfo{volume}{108}},
  \bibinfo{pages}{170403} (\bibinfo{year}{2012}).

\end{thebibliography}

\end{document}